\documentclass[algorithms,article,submit,moreauthors,pdftex]{Definitions/mdpi}
\preto{\abstractkeywords}{\nolinenumbers}
\firstpage{1}
\pubvolume{xx}
\issuenum{1}
\articlenumber{5}
\pubyear{2019}
\copyrightyear{2019}
\history{Received: date; Accepted: date; Published: date}

\usepackage{amsfonts,bm}
\usepackage{amsmath}
\usepackage{algorithm}
\usepackage{algpseudocode}
\usepackage{booktabs, caption, siunitx}
\usepackage{comment}
\usepackage{csquotes}
\usepackage{placeins}
\usepackage[colorinlistoftodos,prependcaption,textsize=scriptsize]{todonotes}
\usepackage{framed}
\usepackage{xspace}
\usepackage[normalem]{ulem}
\usepackage{url}
\usepackage{tikz}
\usetikzlibrary{arrows.meta}
\usepackage{pgfplots}
\pgfplotsset{compat=1.13}
\usepackage[T1]{fontenc}
\usepackage{lmodern}
\usepackage{numprint}
\npdecimalsign{.}
\usepackage{wrapfig}
\usepackage{subcaption}
\usepackage{verbatimbox}

\definecolor{blue1}{RGB}{44,127,184}
\definecolor{shadecolor}{rgb}{0.85,0.85,0.85}
\definecolor{blue2}{RGB}{10,50,250}
\definecolor{markedcolor}{RGB}{31,120,180}
\definecolor{plottinggreen}{RGB}{178,223,138}

\graphicspath{{./images/}}

\usepackage{xspace}

\newcommand{\ie}{i.\,e.,\xspace}
\newcommand{\eg}{e.\,g.,\xspace}
\newcommand{\etc}{etc.\xspace}

\newcommand{\etal}{et al.\xspace}

\newcommand{\quot}[1]{``#1''}

\newcommand{\boldTildex}{{\ensuremath{\widetilde{\bm{x}}}}}
\newcommand{\boldTildeB}{{\ensuremath{\widetilde{\bm{b}}}}}

\newcommand{\mathbc}{{\ensuremath\bm{b}}}
\renewcommand{\phi}{\varphi}

\newcommand{\Oh}{{\ensuremath{\mathcal{O}}}}
\newcommand{\NP}{{\ensuremath{\mathcal{NP}}}}

\newcommand{\tool}[1]{\textsf{#1}}
\newcommand{\nwk}{\tool{NetworKit}\xspace}
\newcommand{\kad}{\tool{KADABRA}\xspace}
\newcommand{\rk}{\tool{RK}\xspace}
\newcommand{\exptool}{\tool{SimexPal}\xspace}
\newcommand{\expcmd}{\texttt{simex}\xspace}

\Title{Guidelines for Experimental Algorithmics in Network Analysis}

\Author{Eugenio Angriman$^{1,\ddagger}$, Alexander van der Grinten$^{1,\ddagger}\orcidC{}$, Moritz von Looz$^{1,\ddagger}$, Henning Meyerhenke$^{1,\ddagger,*}$\orcidA{},
Martin N\"ollenburg$^{2,\ddagger}$\orcidB{}, Maria Predari$^{1,\ddagger}$ and Charilaos Tzovas$^{1,\ddagger}$}

\AuthorNames{Eugenio Angriman, Alexander van der Grinten, Moritz von Looz,
Martin N\"ollenburg, Maria Predari, Charilaos Tzovas and Henning Meyerhenke}

\address{%
$^{1}$ \quad Department of Computer Science, Humboldt-Universität zu Berlin, Germany; \{angrimae, avdgrinten, loozmori, meyerhenke, predarim, charilat\}@hu-berlin.de\\
$^{2}$ \quad Institute of Logic and Computation, Vienna University of Technology, Austria; noellenburg@ac.tuwien.ac.at}

\corres{Correspondence: meyerhenke@hu-berlin.de}
\secondnote{These authors contributed equally to this work.}

\abstract{
The field of network science is a highly interdisciplinary area; for the empirical analysis of network data,
it draws algorithmic methodologies from several research fields. Hence, research procedures and descriptions 
of the technical results often differ, sometimes widely.\\
In this paper we focus on \emph{methodologies} for the experimental part of algorithm engineering for network 
analysis -- an important ingredient for a research area with empirical focus.
More precisely, we unify and adapt existing recommendations from different
fields and propose universal guidelines -- including statistical analyses -- for the systematic evaluation
of network analysis algorithms. This way, the behavior of newly proposed algorithms can be properly assessed and comparisons to existing solutions become meaningful.
Moreover, as the main technical contribution,
we provide \exptool, a highly automated tool to perform and analyze experiments following our guidelines.
To illustrate the merits of \exptool and our guidelines, we apply them in a case study: 
we design, perform, visualize and evaluate experiments of a recent algorithm for 
approximating betweenness centrality, an important problem in network analysis.\\ 
In summary, both our guidelines and \exptool shall modernize and complement previous efforts in experimental
algorithmics; they are not only useful for network analysis, but also in related contexts.
}

\keyword{Experimental algorithmics, network analysis, applied graph algorithms, statistical analysis of algorithms}

\begin{document}
\maketitle
\section{Introduction}
\label{sec:intro}
The traditional algorithm development process in theoretical computer science typically involves
(i) algorithm design based on abstract and simplified models and
(ii) analyzing the algorithm's behavior within these models using analytical techniques.
This usually leads to asymptotic results, mostly regarding the worst-case performance of the algorithm.
(While average-case~\cite{TCS-004} and smoothed analysis~\cite{Spielman:2001:SAA:380752.380813} exist and gained some popularity,
worst-case bounds still make up the vast majority of
running time results.)
Such worst-case results are, however, not necessarily representative for algorithmic behavior in real-world situations, both for \NP-complete problems~\cite{HeuleJS18sat,Applegate:2007:TSP:1374811} and poly-time ones~\cite{mehlhorn2008algorithms,Puglisi:2007:TSA:1242471.1242472}.
In case of such a discrepancy, deciding upon the best-fitted algorithm solely based on worst-case bounds is ill-advised.

\emph{Algorithm engineering} has been established to overcome such pitfalls~\cite{Johnson99,Muller-Hannemann10,Moret}.
In essence, algorithm engineering is a cyclic process that consists of five iterative phases:
(i) modeling the problem (which usually stems from real-world applications),
(ii) designing an algorithm, (iii) analyzing it theoretically, (iv) implementing it, and
(v) evaluating it via systematic experiments (also known as \emph{experimental algorithmics}).
Note that not all phases have to be reiterated necessarily in every cycle~\cite{Sanders10}. 
This cyclic approach aims at a symbiosis: the experimental results shall yield insights that lead to further theoretical improvements and vice versa.
Ideally, algorithm engineering results in algorithms that are asymptotically optimal \emph{and} have excellent behavior in practice at the same time.
Numerous examples where surprisingly large improvements could be made through algorithm engineering exist,
\eg routing in road networks~\cite{bast2016route} and mathematical programming~\cite{applegate2006traveling}.

In this paper, we investigate and provide guidance on the experimental algorithmics part of algorithm
engineering -- from a \emph{network analysis} viewpoint.
It seems instructive to view network analysis, a subfield of \emph{network science}, from two perspectives:
on the one hand, it is a collection of methods that study the structural and algorithmic aspects of networks
(and how these aspects affect the underlying application). The research focus here is on efficient
algorithmic methods.
On the other hand, network analysis can be the process of interpreting network data using the above methods.
We briefly touch upon the latter; yet, this paper's focus is on experimental evaluation methodology,
in particular regarding the underlying (graph) algorithms developed as part of the network analysis toolbox.\footnote{We use the terms \emph{network} and \emph{graph} interchangeably.}

In this view, network analysis constitutes a subarea of empirical graph algorithmics and statistical analysis
(with the curtailment that networks constitute a particular data type)~\cite{brandes_robins_mccranie_wasserman_2013}.
This implies that, like general statistics, network science is not tied to any particular application area.
Indeed, since networks are abstract models of various real-world phenomena, network science draws applications from very diverse
scientific fields such as social science, physics, biology and computer science~\cite{newman2018networks}.
It is interdisciplinary and all fields at the interface have their own expertise and methodology.
The description of algorithms and their theoretical and experimental analysis often differ, sometimes widely -- depending on the target community.
We believe that uniform guidelines would help with respect to comparability and systematic presentation.
That is why we consider our work (although it has limited algorithmic novelty) important for the field 
of network analysis (as well as network science and empirical algorithmics in general). 
After all, emerging scientific fields should develop their own best practices.

To stay focused, we concentrate on providing \emph{guidelines for the experimental algorithmics} part
of the algorithm engineering cycle -- with emphasis on graph algorithms for network analysis.
To this end, we combine existing recommendations 
from fields such as  statistical analysis and data mining / machine learning and 
adjust them to fit the idiosyncrasies of networks.
Furthermore, and as main technical contribution, we provide \exptool, 
a highly automated tool to perform and analyze experiments following our guidelines.
For illustration purposes, we use this tool in a case study --
the experimental evaluation of a recent algorithm for approximating \emph{betweenness centrality},
a well-known network analysis task.
The target audience we envision consists of network analysts who develop algorithms and evaluate them
empirically. Experienced undergraduates and young PhD students will probably benefit most,
but even experienced researchers in the community may benefit substantially from \exptool.

\section{Common Pitfalls (and How to Avoid Them)}
\label{sec:pitfalls-outline}
\subsection{Common Pitfalls}
\label{sub:pitfalls}
Let us first of all consider a few pitfalls to avoid.\footnote{Such pitfalls are probably more often on
a reviewer's desk than one might think.
Note that we do not claim that we never stumbled into such pitfalls ourselves, nor that we
always followed our guidelines in the past.
}
We leave problems in modeling the underlying real-world problem aside and
focus on the evaluation of the algorithmic method for the problem at hand.
Instead, we discuss a few examples from two main categories of pitfalls: (i) inadequate justification of claims on
the paper's results and (ii) repeatability/replicability/reproducibility\footnote{
The terms will be explained below; for more detailed context please visit ACM's webpage on
artifact review and badging: \url{https://www.acm.org/publications/policies/artifact-review-badging}.}
issues.

Clearly, a paper's claims regarding the algorithm's empirical behavior need an adequate justification
by the experimental results and/or their presentation.
Issues in this regard may happen for a variety of reasons; not uncommon is a lack of instances
in terms of their number, variety and/or size. 
An inadequate literature search may lead to not comparing against the current state of the art or if doing so, choosing unsuitable parameters.
Also noteworthy is an inappropriate usage of statistics.
For example, arithmetic averages over a large set of instances might be skewed towards the more
difficult instances. Reporting only averages would not be sufficient then.
Similarly, if the difference between two algorithms is small, it is hard to decide
whether one of the algorithms truly performs better than the other one
or if the perceived difference is only due to noise.
Even if no outright mistakes are made, potential significance can be wasted:
Coffin and Saltzmann~\cite{Coffin00}
discuss papers whose claims could have been strengthened 
by an appropriate statistical analysis,
even without gathering new experimental data.

\emph{Reproducibility} (the algorithm is implemented and evaluated independently by a different team)
is a major cornerstone of science and should receive sufficient attention.
Weaker notions include \emph{replicability} (different team, same source code and experimental setup)
and \emph{repeatability} (same team, same source code and experimental setup).
An important weakness of a paper would thus be if the description does not
allow the (independent) reproduction of the results.
First of all, this means that the proposed algorithm
needs to be explained adequately. 
If the source code is not public, it is all the more
important to explain all important implementation aspects
-- the reader may judge whether this is the current community standard.
Pseudocode allows to reason about correctness and asymptotic time and space complexity and is thus
very important.
But the empirical running time of a reimplementation
may deviate from the expectation when the paper omitted
crucial implementation details.

Even the repetition of one's own results (which is mandated by major funding organizations for time spans such as 10 years)
can become cumbersome if not considered from the beginning.
To make this task easier, not only source code needs to documented properly,
but also input and output data of experiments as well as the scripts containing the parameters.
Probably every reader has heard about this one project where this
documentation part has been neglected to some extent...

\subsection{Outline of the Paper}
\label{par:outline}
How do we avoid such pitfalls? We make this clear by means of a use case featuring betweenness approximation;
it is detailed in Section~\ref{sec:target-use-cases}.
A good start when designing experiments is to formulate a hypothesis (or several ones) on the algorithm's behavior 
(see Section~\ref{sub:guidelines:hypotheses}).
This approach is not only part of the scientific method;
 it also helps in structuring the experimental design
and evaluation, thereby decreasing the likelihood of some pitfalls.
Section~\ref{sub:instances} deals with how to select and describe input instances to support a certain
generality of the algorithmic result.
With a similar motivation, Section~\ref{sub:variance} provides guidance on \emph{how many} instances should be selected.
It also discusses how to deal with variance in case of non-determinism, \eg by stating how 
often experiments should be repeated.

If the algorithm contains a reasonable amount of tunable parameters,
a division of the instances into a tuning set and an evaluation set may be advisable,
which is discussed in Section~\ref{sub:training-test-set}.

In order to compare against the best competitors, one must have defined which algorithms and/or
software tools constitute the state of the art. Important aspects of this procedure are discussed in Section~\ref{sub:competition}. One aspect can be the approximation quality:
while an algorithm may have the better quality guarantee in theory, its empirical
solution quality may be worse -- which justifies to consider not only the theoretical reference in experiments.
The claim of superiority refers to certain measures -- typically related
to resources such as running time or memory and/or related to solution
quality. Several common measures of this kind and how to deal with them are explained in Section~\ref{sub:metrics}.
Some of them are hardware-independent, many of them are not.

A good experimental design can take you far -- but it is not sufficient if the experimental
pipeline is not efficient or lacks clarity. In the first case, obtaining the results may be tedious
and time-consuming.
Or the experiments simply consume more computing resources than
necessary and take too long to generate the results. In the second case, in turn, the way experimental
data are generated or stored may not allow easy reproducibility. 
Guidelines on how to setup your experimental pipeline and how to avoid these pitfalls are thus presented
in Section~\ref{sec:exp-pipeline}. The respective subsections deal with implementation 
issues~(\ref{sub:implementation}), repeatability/replicability/reproducibility (\ref{sub:reproducibility}), job submission 
(\ref{sub:submission-collection}), output file structure for long-term storage (\ref{sub:benchmark-output}),
retrieval and aggregation of results~(\ref{sub:gather-results}).

As mentioned, betweenness approximation and more precisely the \kad~\cite{borassi2016kadabra}
algorithm will serve as our prime example.
We have implemented this algorithm as part of the \nwk toolbox~\cite{staudt2016networkit}.
As part of our experimental evaluation, a meaningful visualization (Section~\ref{sec:viz}) of the results
highlights many major outcomes to the human reader.
Since visualization is typically not enough to show statistical significance in a rigorous manner,
an appropriate statistical analysis (Section~\ref{sec:statistical-analysis}) is recommended.
Both, visualization and statistical analysis, shall lead to a justified interpretation of the results.

\section{Use Case}
\label{sec:target-use-cases}
Typically, algorithmic network analysis papers (i) contribute 
a new (graph) algorithm for an already known problem that improves on the
state of the art in some respect or
(ii) they define a new algorithmic problem and present a method for its solution.
To consider a concrete example contribution, we turn towards {betweenness centrality},
a widely used and very popular measure for ranking nodes (or edges) in terms of their structural position in
the graph~\cite{boldi2014axioms}.
\subsection{Betweenness Approximation as Concrete Example}
\label{sub:concrete-example}
Betweenness centrality~\cite{freeman1977set} measures the participation of nodes in 
shortest paths. More precisely, let $G = (V,E)$ be a graph; the (normalized) betweenness centrality of a node $v \in V$ is defined as
\begin{equation}
\label{eq:betweenness_def}
  \mathbc(v) := \frac{1}{n (n-1)} \sum_{s, t \in V, s\neq v \neq t}{\frac{\sigma_{st}(v)}{\sigma_{st}}},
\end{equation}
where $\sigma_{st}$ is the number of shortest paths from $s$ to $t$, and $\sigma_{st}(v)$ is the number of shortest paths from $s$ to $t$ that cross $v$ (as intermediate node).
Computing the exact betweenness scores for all nodes of an unweighted graph can be done in $\Oh(nm)$ time
with Brandes's algorithm~\cite{brandes2001faster}, where $n$ is the number of nodes and $m$ the number of edges in $G$.
Since this complexity is usually too high for graphs with millions of nodes and edges, 
several approximation algorithms have been 
devised~\cite{bader2007approximating,geisberger08,riondato2016fast,riondato2018abra,borassi2016kadabra}.
These algorithms trade solution quality for speed and can be much faster.

For illustration purposes, we put ourselves now in the 
shoes of the authors of the most recent of the cited algorithms, which is called \kad~\cite{borassi2016kadabra}:
we describe (some of) the necessary steps in the process of writing an algorithm engineering
paper on \kad\footnote{We select \kad not because the authors of the original paper did a bad job
in their presentation -- \textit{au contraire}. Our main reasons are (i) the high likelihood that the betweenness problem
is already known to the reader due to its popularity in the field and (ii) the fact that approximation
algorithms display important aspects of experimental algorithmics quite clearly.} --
with a focus on the design and evaluation of the experiments.

\subsection{Overview of the \kad Algorithm}
\label{sub:kadabra}
\begin{algorithm}[bt]
\caption{\kad algorithm (absolute error variant)}
\label{algo:kadabra}
\setlength\baselineskip{16pt}
\begin{algorithmic}[1]
\Procedure{\kad}{$G = (V, E)$, $\epsilon$, $\delta$}
	\State $\omega \gets$ non-adaptive number of iterations
	\State compute $\delta_L, \delta_U$ from $\delta$ \label{line:delta_guess}
		\Comment{Requires upper bound of the diameter.}
	\State $\tau \gets 0$ \Comment{Number of sampled paths.}
	\ForAll{$v \in V$}
		\State $\boldTildex(v) \gets 0$ \Comment{Occurrences of $v$ in sampled paths.}
	\EndFor
	\While{$\tau < \omega$}
		\If{$f(\boldTildex(v)/\tau,~\delta_L(v),~\omega,~\tau) < \epsilon$ \textbf{and}
				$g(\boldTildex(v)/\tau,~\delta_U(v),~\omega,~\tau) < \epsilon$}
				\label{line:adaptive}
			\State \textbf{break} \Comment{Approximation is already good enough.}
		\EndIf
		\For{$i \in \{1, \ldots, c\}$} \Comment{Draw $c$ samples per round.}
			\State $s, t \gets$ \texttt{samplePair}($G$)
					\Comment{Uniformly at random.}
			\State $\pi \gets$ \texttt{sampleShortestPath}($G, s, t$)
					\Comment{Uniformly at random.}
			\ForAll{$v \in \pi$}
				\State $\boldTildex(v) \gets \boldTildex(v) + 1$
			\EndFor
			\State $\tau \gets \tau + 1$
		\EndFor
	\EndWhile
	\State \Return $\boldTildex / \tau$
			\Comment{Betweenness centrality is estimated as: $\boldTildeB = \boldTildex/\tau$.}
\EndProcedure
\end{algorithmic}
\end{algorithm}
The \kad algorithm approximates the betweenness centrality of
(un)directed graphs within a given absolute error of at most
$\epsilon$ with probability
$(1 - \delta)$~\cite{borassi2016kadabra}.
The main rationale of the algorithm is to iteratively select two nodes $s$, $t$ uniformly at
random and then sample a shortest path $\pi$
from $s$ to $t$ (again uniformly at random).
This leads to a sequence of randomly selected shortest paths
$\pi_1, \pi_2, \dots, \pi_\tau$.
The betweenness centrality of each node $v \in V$ is then estimated as:
\begin{equation}
\label{eq:betweenness_estimation}
  \boldTildeB(v) = \frac{1}{\tau}\sum_{i = 1}^{\tau}{\boldTildex_i(v)},
\end{equation}

where $\boldTildex_i$ is 1 iff $v$ occurs in $\pi_i$ and 0 otherwise.

Compared to earlier approximation algorithms that employ
similar sampling techniques (\eg \cite{riondato2016fast}),
the novelty of \kad relies on the clever stopping condition, used to determine
the number of rounds $\tau$. Clearly, there is a number $\omega$,
depending on the input size but not the graph itself,
such that if $\tau \geq \omega$, the algorithm achieves the desired solution quality
with probability $(1 - \delta)$.\footnote{For example,
	if $\omega$ is chosen so that almost all vertex pairs are sampled.
	In reality, $\omega$ can be chosen to be much smaller than that.}
\kad, however, avoids to run a fixed number of $\omega$ rounds by using
\emph{adaptive sampling}. At each round of the algorithm, it is guaranteed that
\begin{equation}
  \textnormal{Pr}\left( \mathbc(v) \le \boldTildeB(v) - f \right) \le \delta_L(v)
  \quad \textnormal{and} \quad
  \textnormal{Pr}\left( \mathbc(v) \ge \boldTildeB(v) + g \right) \le \delta_U(v),
\end{equation}
where $f = f(\boldTildeB(v), \delta_L(v), \omega, \tau)$
and $g = g(\boldTildeB(v), \delta_U(v), \omega, \tau)$
are (rather lengthy) expressions depending on
$\boldTildeB(v)$, per-vertex probabilities
$\delta_L(v)$ and $\delta_U(v)$,
the current number of rounds $\tau$ and the static round count
$\omega$.
$\delta_L(v)$ and $\delta_U(v)$ are chosen such that
$\sum_{v \in V} \delta_L(v) + \delta_U(v) \leq \frac\delta2$.
Once $f, g < \epsilon$ during some round,
the algorithm terminates.\footnote{
Note also that for each round,
  the algorithm draws a number of samples and performs occurrence updates
  without checking the stopping condition.
  This number is determined by parameter $c$. In the original
  implementation of \kad (but not reported in the paper), $c$ is fixed to $10$ (without further explanation).}
Algorithm~\ref{algo:kadabra} displays the corresponding pseudocode with some
expository comments.
Besides adaptive sampling, \kad
relies on a \emph{balanced bidirectional BFS} to sample shortest paths.
For details we refer the interested reader to the original paper.

\section{Guidelines for the Experimental Design}
Now, we want to set up experiments that give decisive and convincing empirical evidence whether \kad is better than the state of the art.
This section discusses the most common individual steps of this process.

\subsection{Determining Your Hypotheses}
\label{sub:guidelines:hypotheses}
Experiments are mainly used for two reasons;
as an exploratory tool to reveal unknown properties
and/or as a means to answer specific questions regarding the proposed algorithm.
The latter suggests the development of certain hypotheses
on the behavior of our algorithm that can be confirmed or contradicted by experiments.
In the following, we group common hypotheses of algorithm
engineering papers in network 
analysis into two categories:

\begin{enumerate}
\item \label{hyp:1} Hypotheses on how our algorithm performs compared to
  the state of the art in one of the 
  following metrics: running time, solution quality or
  (less often) memory usage.
  Ideally, an algorithm is deemed successful when it outperforms
  existing algorithms in terms of all
  three metrics. However, in practice,
  we are content with algorithms that exhibit a good tradeoff 
  between two metrics, often running time performance and
  solution quality.
  As an example,
  in real-time applications, 
  we may be willing to sacrifice the solution quality
  (until a certain threshold), in order to meet
  running time bounds crucial for the application.

\item \label{hyp:2} Hypotheses on how the input instances or a
  problem-/algorithm-specific parameter affects our algorithm in terms of the
  aforementioned metrics. 
  For example, a new algorithm might only outperform the
  existing algorithms on a certain type of graphs or an
  approximation algorithm might only be fast(er) if 
  the desired approximation quality is within a certain range.
  If a hypothesis involves such
  restrictions, it should still be general enough for the algorithmic result to be relevant --
  overtuning of algorithms to uninteresting classes of
  inputs should be avoided.
  Other aspects of investigation may be decoupled from the state of the art to some
  extent and explore the connection between instance and algorithm properties:
  for instance, a hypothesis on how the empirical approximation error behaves over
  a range of input sizes.

\end{enumerate}

\paragraph{\kad Example}
In the context of our betweenness approximation study,
  we formulate three basic hypotheses:
  \begin{enumerate}
  \item \kad has a better running time and scalability (with respect to the graph size)
  	than other algorithms, specifically the main competitor \rk~\cite{riondato2016fast}.
  \item There is a significant difference between the solution quality of \kad and that of \rk.
   (In Section~\ref{sub:metrics}, we explain how to evaluate the solution quality for approximation algorithms in more detail.)
  \item The diameter of input graphs has an effect on the running time of \kad:
   The \kad algorithm computes the betweenness values faster for graphs with low diameter than for graphs with large diameter.
  \end{enumerate}

  The first two hypotheses belong to the first category, since we compare
  our implementation of the \kad algorithm to a number of related algorithms.
The other hypothesis belongs to the second category and is related to
the performance of the \kad algorithm itself. Namely,
we test how a data-specific parameter
may influence the running time in practice.
To evaluate these hypotheses experimentally, we need to select
instances with both low and high diameter values, of course.

\subsection{Gathering Instances}
\label{sub:instances}
Selecting appropriate data sets for an experimental evaluation is a crucial design step.
For sake of comparability, a general rule is to gather data sets from public sources;
for network analysis purposes, well-known repositories are KONECT~\cite{kunegis2013konect}, SNAP~\cite{snapnets}, DIMACS10~\cite{dimacs},
SuiteSparse~\cite{davis2011university}, LWA~\cite{BoVWFI} and
Network Repository~\cite{nr-sigkdd16}.

An appropriate data collection contains the type(s) of networks which the algorithm is designed for, and,
for every type, a large enough number of instances to support our conclusions~\cite{Johnson99}.
For instance, the data collection of an influence maximization algorithm should include social networks rather than
street networks; a data collection to evaluate distributed algorithms
should include instances that exceed the memory of a typical shared-memory machine.

The selection of an appropriate subset of instances for our experimental analysis is simpler if we first categorize the available networks in different classes.
There exist several ways to do this:
first, we can simply distinguish real-world and synthetic (= (randomly) generated) networks.
Even though network analysis algorithms generally target real-world networks, one should also use synthetic 
networks, \eg to study the asymptotic scalability of an algorithm, since we can easily generate similar synthetic
networks of different scales.

Classification of real-world networks generally follows the phenomena they are modeling.
Common examples are social networks, hyperlink networks or citation networks, which also fall under the umbrella of \emph{complex networks}.\footnote{As the name suggests, complex networks have highly non-trivial (complex) topological features.
Most notably, such features include a small diameter (\emph{small-world effect}) and a skewed degree distribution
(many vertices with low degree and a few vertices with high degree).
}
Examples of real-world non-complex networks are certain infrastructure networks such as road networks.
If our algorithm targets real-world data, we want to carefully build a diverse collection of instances that is 
representative of the graph classes we are targeting.

Another interesting classification is one based on specific topological features of the input data
such as the diameter, the clustering coefficient, the triangle count, etc. Classifying based on a certain
property and reporting it should definitely be done when the property in question is relevant to the
algorithm and may have an impact on its behavior. For instance,
reporting the clustering coefficient could help with the interpretation of results
concerning algorithms that perform triangle counting.

A source of instances especially suited for scalability experiments are synthetic graphs generated with graph generators~\cite{goldenberg2010survey}.
An advantage is easier handling especially for large graphs, since no large files need to be transferred and stored.
Also, desired properties can often be specified in advance and in some models, a ground truth beneficial to test analysis algorithms is available.
Important drawbacks include a lack of realism, especially when choosing an unsuitable generative model.

\begin{table}[tb]
\caption{Graphs (taken from KONECT) for our evaluation of the betweenness approximation algorithms.}
\label{tab:kadabra_dataset}
\centering
\footnotesize
\begin{tabular}{lrrrr}
\toprule
Network name & \# of nodes & \# of edges & Diameter & Class\\
\midrule
\midrule
moreno\_blogs & \numprint{1224} &
			\numprint{16715} &
			\numprint{8} &
			Hyperlink \\
petster-hamster & \numprint{2426} &
			\numprint{16631} &
			\numprint{10} &
			Social \\
ego-facebook & \numprint{2888} &
			\numprint{2981} &
			\numprint{9} &
			Social \\
openflights & \numprint{3425} &
			\numprint{19256} &
			\numprint{13} &
			Infrastructure \\
opsahl-powergrid & \numprint{4941} &
			\numprint{6594} &
			\numprint{46} &
			Infrastructure \\
p2p-Gnutella08 & \numprint{6301} &
			\numprint{20777} &
			\numprint{9} &
			Peer-to-peer \\
advogato & \numprint{6539} &
			\numprint{39285} &
			\numprint{9} &
			Social \\
wiki-Vote & \numprint{7115} &
			\numprint{100762} &
			\numprint{7} &
			Social \\
p2p-Gnutella05 & \numprint{8846} &
			\numprint{31839} &
			\numprint{9} &
			Peer-to-peer \\
p2p-Gnutella04 & \numprint{10876} &
			\numprint{39994} &
			\numprint{10} &
			Peer-to-peer \\
foldoc & \numprint{13356} &
			\numprint{91471} &
			\numprint{8} &
			Hyperlink \\
twin & \numprint{14274} &
			\numprint{20573} &
			\numprint{25} &
			Intl. Relations \\
cfinder-google & \numprint{15763} &
			\numprint{148585} &
			\numprint{7} &
			Hyperlink \\
ca-AstroPh & \numprint{18771} &
			\numprint{198050} &
			\numprint{14} &
			Coauthorship \\
ca-cit-HepTh & \numprint{22908} &
			\numprint{2444798} &
			\numprint{9} &
			Coauthorship \\
subelj\_cora & \numprint{23166} &
			\numprint{89157} &
			\numprint{20} &
			Citation \\
ego-twitter & \numprint{23370} &
			\numprint{32831} &
			\numprint{15} &
			Social \\
ego-gplus & \numprint{23628} &
			\numprint{39194} &
			\numprint{8} &
			Social \\
p2p-Gnutella24 & \numprint{26518} &
			\numprint{65369} &
			\numprint{11} &
			Peer-to-peer \\
ca-cit-HepPh & \numprint{28093} &
			\numprint{3148447} &
			\numprint{9} &
			Citation \\
cit-HepPh & \numprint{34546} &
			\numprint{420877} &
			\numprint{14} &
			Citation \\
facebook-wosn-wall & \numprint{46952} &
			\numprint{183412} &
			\numprint{18} &
			Social \\
edit-frwikibooks & \numprint{47905} &
			\numprint{139141} &
			\numprint{8} &
			Authorship \\
dblp-cite & \numprint{49789} &
			\numprint{49759} &
			\numprint{2} &
			Citation \\
loc-brightkite\_edges & \numprint{58228} &
			\numprint{214078} &
			\numprint{18} &
			Social \\
edit-frwikinews & \numprint{59546} &
			\numprint{157970} &
			\numprint{7} &
			Authorship \\
dimacs9-BAY & \numprint{321270} &
			\numprint{397415} &
			\numprint{837} &
			Road \\
dimacs9-COL & \numprint{435666} &
			\numprint{521200} &
			\numprint{1255} &
			Road \\
roadNet-PA & \numprint{1088092} &
			\numprint{1541898} &
			\numprint{794} &
			Road \\
roadNet-TX & \numprint{1379917} &
			\numprint{1921660} &
			\numprint{1064} &
			Road \\
\midrule
\midrule
\end{tabular}

\end{table}

\paragraph{\kad Example}
In a real-world scenario, we want to show the improvement of the \kad algorithm with respect to the competition.
Clearly, the graph class of highest relevance should be most prominent in our data collection.
With this objective in mind, we show a collection for the \kad experiments in Table~\ref{tab:kadabra_dataset}.
Note that in Table~\ref{tab:kadabra_dataset}
we report the diameter of each graph, along with its size and application class.
We include the diameter because it is part of our hypotheses, \ie
the performance of \kad depends on the diameter of the input graph
(see Section~\ref{sub:guidelines:hypotheses}).
In a different scenario, such as triangle counting or community detection, a more pertinent choice would
be the average local clustering coefficient.
Finally, we focus on complex networks, but include a minority of non-complex infrastructure networks.
All of these instances were gathered from one of the aforementioned public repositories.

\subsection{Scale of Experiments}
\label{sub:variance}
Clearly, we cannot test all possible instances, even if they are small~\cite{McGeochP00}.
On the other hand, we want to draw significant conclusions about our new algorithm (and here not about
a particular data set). This means that our experimental results should justify the later conclusions 
and allow to clearly distinguish between hypotheses.
While Section~\ref{sub:instances} discusses \emph{which} instances to select, we proceed with the question
\emph{``how many?''}.
Also, if the results are affected by randomness, how many repeated runs should we make for each?

Too few instances may be insufficient to support a conclusion.
Plots then look sparse, hypothesis tests (Section~\ref{sec:stats:nhst}) are inconclusive
and inferred parameters are highly uncertain, which is reflected in unusably wide confidence and credible intervals
(Sections~\ref{sec:stats:confidence-intervals} and \ref{sec:stats:bayesian-inference}).
Choosing too many instances, though, costs unnecessary time and expense.
As a very rough rule of thumb reflecting community custom, we recommend at least 10-15 instances for an experimental paper.
If you want to support specific conclusions about algorithmic behavior within a subclass of instances, 
you should have that many instances \emph{in that subclass}.
Yet, the noisier the measurements and the stronger the differences of output between instances, the more experiments are needed for a meaningful (statistical) analysis.
More formally, the uncertainty of many test statistics and inferred parameters\footnote{Most importantly the t-statistic used in confidence intervals and t-tests.
 In Bayesian statistics, the marginal posterior distribution of a parameter with a Gaussian likelihood also follows a t-distribution, at least when using a conjugate prior~\cite{bernardo2009bayesian}.
 } scales with $\sqrt{\frac{s^2}{n}}$, where $s^2$ is the sample variance.
 Generally speaking, if we want a result that is twice as precise, we need to quadruple the number of measurements.

The same applies for the number of repetitions for each instance.
When plotting results affected by randomness, it is often recommended to average over the results of multiple runs.
The expected deviation of such an average from the true mean\footnote{The mean that would result given infinite samples.} is called the \emph{standard error}
and also scales with $\frac{s^2}{\sqrt{n}}$~\cite{Altman05}.
For plotting purposes, we suggest that this standard error for the value of one plotted point (\ie the size of the error bars) should be one order of magnitude smaller than the variance \emph{between} plotted points.
In the companion notebook, we give an example of code which calculates the number of repetitions required for a desired smoothness.
Frameworks exist that automate the repeated measurements, for example Google Benchmark.\footnote{\url{https://github.com/google/benchmark}.}

In the context of binary hypothesis tests, the \emph{statistical power} of a test is the probability that it correctly distinguishes between hypotheses,
\ie rejects a false null hypothesis in the classical approach.
A larger sample size, a stronger effect or less variance in measurements all lead to a more powerful test.
For further reading, see the extensive literature about statistical power~\cite{ellis2010essential,Cohen88}.

All these calculations for the number of necessary experiments require an estimate of the variance of results -- which might be unavailable before conducting the experiments.
It might thus be necessary to start with a small initial set of instances to plan and prepare for the main experiments.\footnote{Also called \emph{pilot study}, as opposed to the main experiments called \emph{workhorse study}, see~\cite{McGeoch12} and~\cite{Rardin2001}.}

\subsection{Parameter Tuning}
\label{sub:training-test-set}

Experiments are not only a necessary ingredient to validate our hypothesis, but also a
helpful discovery tool in a previous stage~\cite{Moret}.
Some algorithm's behavior depends on tunable parameters.
For example, the well-known hybrid implementation of Quicksort switches between Insertionsort
and Quicksort depending on whether the current array size is above or below a tunable threshold
$M$~\cite{sedgewick1978implementing}.
An example from network analysis includes a plethora of approximation methods
that calculate some centrality measure using
sampling~\cite{geisberger08,Eppstein:2001:FAC:365411.365449,riondato2016fast}.
In such cases the number of selected samples
may highly impact the performance of the algorithm.
Experiments are thus used to determine the adequate number of samples that achieves a good approximation.
  
The \kad algorithm consists of two phases, in which bounds for a sampling
process are first computed (Line~\ref{line:delta_guess} in
Algorithm~\ref{algo:kadabra} of Section~\ref{sub:kadabra})
and then used (when checking the stopping condition in line~\ref{line:adaptive}).
Tighter bounds are more expensive to compute but save time in the later phase;
their quality is thus an important tradeoff.
When evaluating a newly developed algorithm, there is the risk (or even temptation) to tune its parameters for optimal performance in the experiments while using general default parameters for all competitors.
This may often happen with no ill intent -- researchers usually know their own algorithm best.
To ensure generalizability of results, we recommend to split the set of instances into a \emph{tuning set} and an \emph{evaluation set}.
The tuning set is used to find good tuning parameters for all algorithms, while the main experiments are performed on the evaluation set.
Only results from the evaluation set should be used to draw and support conclusions.
Note that the tuning set can also be used to derive an initial variance estimate, as discussed in Section~\ref{sub:variance}.

\paragraph{How to Create the Tuning Set and the Evaluation Set}
The tuning set should be structurally similar to the whole data set, so that parameter tuning yields good general performance and is representative for the evaluation set.\footnote{Due to symmetry, this also requires that the tuning set is structurally similar to the evaluation set.}
Tuning and evaluation sets should be disjoint.
For large data sets, simple random sampling (\ie simply picking instances from the data set uniformly at random and without replacement) yields such a representative split in expectation.
Stratified sampling (\ie partitioning the data set, applying simple random sampling to each partition and merging the result)~\cite{tibshirani2013introduction} guarantees it.

In our example, note that our data set is already partitioned into different classes of networks (hyperlink, social, infrastructure, \etc), and that those classes are fairly balanced (see~Table~\ref{tab:class_freq}).
We can thus select a certain fraction of instances in each network class as tuning set.

In case of computationally expensive tuning, it is advantageous to keep the tuning set small, for large data sets it can be much smaller than the evaluation set.
In our example, we select a single instance per network class, as seen in Table~\ref{tab:kadabra_tuning_set}.
Note that these considerations are similar to the creation of training, test and validation sets in machine learning.
While some other sampling methods are equally applicable, more sophisticated methods like statistical learning theory optimize for different objectives and are thus not considered here~\cite{hinton2012practical, larochelle2007empirical, lecun1998gradient}.

\begin{table}
\centering
\caption{Network class frequency of the data set in Table~\ref{tab:kadabra_dataset}}
\label{tab:class_freq}
\begin{tabular}{lr}
\toprule
Class & Frequency\\
\midrule
\midrule
Social & \numprint{26.67} \%\\
Citation & \numprint{13.33} \%\\
Peer-to-peer & \numprint{13.33} \%\\
Road & \numprint{13.33} \%\\
Hyperlink & \numprint{10.00} \%\\
Infrastructure & \numprint{6.67} \%\\
Coauthorship & \numprint{6.67} \%\\
Authorship & \numprint{6.67} \%\\
Intl. Relationship & \numprint{3.33} \%\\
\midrule
\midrule
\end{tabular}

\end{table}

\begin{table}
\centering
\caption{Split of instances into tuning and evaluation set.}
\label{tab:kadabra_tuning_set}
\footnotesize
\begin{tabular}{@{}l!{\quad}r!{\quad}r!{\quad}l!{\quad}r!{\quad}r@{}}
\toprule
\multicolumn{3}{c}{Tuning set} & \multicolumn{3}{c}{Evaluation set} \\
\cmidrule{1-6} \\
Network name & $|V|$ & $|E|$ & Network name & $|V|$ & $|E|$ \\
\midrule
opsahl-powergrid & \numprint{4941} & \numprint{6594} &moreno\_blogs & \numprint{1224} & \numprint{16718} \\
advogato & \numprint{6539} & \numprint{43277} &  petster-hamster & \numprint{2426} & \numprint{16631}\\
foldoc & \numprint{13356} & \numprint{91471} &  ego-facebook & \numprint{2888} & \numprint{19257} \\
p2p-Gnutella24 & \numprint{26518} & \numprint{65369} & openflights  & \numprint{3425} & \numprint{2981} \\
dblp-cite & \numprint{12591} & \numprint{49635} & p2p-Gnutella08 & \numprint{6301} & \numprint{20777} \\
edit-frwikinews & \numprint{25042} & \numprint{68679}  & wiki-Vote & \numprint{7115} & \numprint{100762} \\
dimacs9-BAY & \numprint{321270} & \numprint{397415} & p2p-Gnutella05 & \numprint{8846} & \numprint{31839} \\ 
&&&p2p-Gnutella04 & \numprint{10876} & \numprint{39994} \\

&&&twin & \numprint{14274} & \numprint{20573} \\
&&&cfinder-google & \numprint{15763} & \numprint{149456} \\
&&&ca-AstroPh & \numprint{18771} & \numprint{198050} \\
&&&ca-cit-HepTh & \numprint{22908} & \numprint{2444798} \\
&&&subelj\_cora & \numprint{23166} & \numprint{89157} \\
&&&ego-twitter & \numprint{23370} & \numprint{32831} \\
&&&ego-gplus & \numprint{23628} & \numprint{39194} \\
&&&edit-frwikibooks & \numprint{27754} & \numprint{67584} \\
&&&ca-cit-HepPh & \numprint{28093} & \numprint{3148447} \\
&&&cit-HepPh & \numprint{34546} & \numprint{420921} \\
&&&loc-brightkite\_edges & \numprint{58228} & \numprint{214078} \\
&&&dimacs9-COL & \numprint{435666} &\numprint{521200} \\
&&&roadNet-PA & \numprint{1088092} &\numprint{1541898}\\
&&&roadNet-TX & \numprint{1379917} &\numprint{1921660} \\

\midrule
\midrule
\end{tabular}
\end{table}

\subsection{Determining Your Competition}
\label{sub:competition}

Competitor algorithms solve the same or a similar problem and are not dominated in the relevant
metrics by other approaches in the literature.
Since we focus on the design of the experimental pipeline, we consider only competitors that
are deemed implementable. The best ones among them are considered the \emph{state of the art} (SotA),
here with respect to the criteria most relevant for our claims about \kad.

Other considerations are discussed next.

\paragraph{Unavailable Source Code} 
The source code of a competing algorithm may be unavailable and sometimes even the executable 
is not shared by its authors.
If one can implement the competing algorithm with reasonable effort 
(to be weighted against the importance of the competitor), you should do so.
If this is not possible, a viable option is to compare experimental results with published data.
For a fair comparison, the experimental settings should then be replicated as closely as possible.
In order to avoid this scenario, we recommend open-source code; it offers better transparency 
and replicability (also see~\ref{sub:reproducibility}).

\paragraph{Solving a Different Problem}
In some cases, the problem our algorithm solves is different from 
established problems and there is no previous algorithm solving it.
Here, it is often still meaningful to compare against the SotA for the established problem.
For example, if we consider the dynamic case of a static problem for the first time 
~\cite{bergamini15,green12,kourtellis14} or an approximation version for an exact problem 
~\cite{brandes07,bader2007approximating, geisberger08}
or provide the first parallel version of an algorithm \cite{bader06}.
Another example can be an optimization problem with a completely new constraint.
While this may change optimal solutions dramatically compared to a previous formulation, a comparison to 
the SotA for the previous formulation may still be the best assessment.
If the problem is so novel that no comparison is meaningful, however, there is no reason
for an experimental comparison.\footnote{Nonetheless, experiments can still reveal empirical insights into the behavior of the new algorithm.}

\paragraph{Comparisons on Different Systems}
Competitors may be designed to run on different systems.
Recently, many algorithms take advantage of accelerators, 
mainly GPUs, and this trend has also affected the algorithmic network analysis community 
\cite{mclaugh14, Sariyuce2013, Shi2011}.
A direct comparison between CPU and GPU implementations is not necessarily meaningful
due to different architectural characteristics and memory sizes.
Yet, such a comparison may be the only choice if no other GPU implementation exists. 
In this case, one should compare against a multithreaded CPU implementation that
should be tuned for high performance as well.
Indirect comparisons can, and should, be done using system independent 
measures (\ref{sub:metrics}).

\paragraph{\kad Example}
The most relevant candidates to compare against 
\kad are the SotA algorithms \rk~\cite{riondato2016fast} and ABRA~\cite{riondato2018abra}.
The same algorithms are chosen as competitors in the original \kad paper. However, in our paper
we focus only on a single comparison, the one between \kad and \rk. This is intended in order
to highlight the main purpose of our work -- to demonstrate the benefits of a thoughtful
experimental design for meaningful comparisons of network-related algorithms.
Reporting the actual results of such a comparison is of secondary importance.
Furthermore, \rk and ABRA exhibit similar behavior in the original \kad paper,
with \rk being overall slightly faster. Again, for the purpose of our paper, this is
an adequate reason to choose \rk over ABRA.\footnote{Of course, if we
  were to perform experiments for a new betweenness approximation algorithm,
  we would include all relevant solutions in the experiments.}

\subsection{Metrics}
\label{sub:metrics}
The most common metric for an algorithm's performance is its wall-clock running time.
For solution quality, the metrics are usually problem-specific;
often, the gap between an algorithm's solution and the optimum is reported
as a percentage (if known).
In the following, we highlight situations that require more specific metrics.
First, we discuss metrics for evaluating an algorithm's \emph{running time}.

\subsubsection{Running Time}

\paragraph*{CPU Time vs. Wall-clock Time}
For sequential algorithms, evaluations should prefer CPU time over wall-clock time.
Indeed, this is the running time metric that we use in our \kad experiments.
Compared to wall-clock time, CPU time is less influenced by external factors
such as OS scheduling or other processes running on the same machine.
The exception to this rule are algorithms that depend on those external factors
such as external memory algorithms (where disregarding the time spent in
system-level I/O routines would be unfair).
In the same line of reasoning, evaluations of parallel algorithms would be
based on wall-clock time as they depend on specifics of the OS scheduler.

\paragraph*{Architecture-specific Metrics}
CPU and wall-clock times heavily depend on the microarchitecture of the CPU
the algorithm is executed on. If comparability with data generated on similar CPUs
from other vendors or microarchitectures is desired, other metrics should be
taken into account. Such metrics can be accessed by utilizing the
\emph{performance monitoring} features of modern CPUs: these allow determining
the number of executed instructions, the number of memory accesses
(both are mostly independent of the CPU microarchitecture)
or the number of cache misses (which depends on the cache size and cache strategy, but not necessarily the CPU model).
If one is interested in how efficient algorithms are in
exploiting the CPU architecture, the utilization of CPU components can be measured
(example: the time spent on executing ALU instructions compared to the time that the ALU is
stalled while waiting for results of previous instructions or memory fetches).
Among some common open-source tools to access CPU performance counters are the \texttt{perf} profiler,
\texttt{Oprofile} (on linux only) and \texttt{CodeXL}.

\paragraph*{System-independent Metrics}
It is desirable to compare different algorithms on the same
system. However, in reality, this is not always possible, \eg because
systems required to run competitors are unavailable. For such cases,
even if we do not run into such issues ourselves, we should also consider
system-independent metrics. For example, this can be the speedup of a newly proposed algorithm against some
base  algorithm that can be tested on all systems.
As an example, for betweenness centrality, some papers re-implement the Brandes~\cite{brandes2001faster}
algorithm and compare their performance against
it~\cite{riondato2018abra, bergamini15, crescenzi15}.
As this metric is independent of the hardware,
it can even be used to compare implementations on different systems, \eg CPU versus GPU implementations.
System-independent metrics also include algorithm-specific metrics,
like the number of iterations of an algorithm or the number of edges visited.
Those are particularly useful when similar algorithms are compared,
\eg if the algorithms are iterative and only differ in their choice of the stopping condition.

\paragraph*{Aggregation and Algorithmic Speedup}
Running time measurements are generally affected by fluctuations, so that the same experiment
is repeated multiple times. To arrive at one value per instance, one usually computes the arithmetic mean
over the experiments (unless the data are highly skewed).
A comparison between the running times (or other metrics) of two algorithms $A$ and $B$ on different
data sets may result in drastically different values, \eg because the instances differ in size or
complexity. For a concise evaluation, one is here also interested in aggregate values.
In such a case it is recommended to aggregate over \emph{ratios} of these metrics;
regarding running time, this would mean to compute the \textit{algorithmic speedup}\footnote{The \emph{parallel speedup} of an algorithm $A$ is instead the speedup of the parallel execution of $A$ against its sequential execution, more precisely the ratio of the running time of the \emph{fastest} sequential algorithm and the running time of the parallel algorithm. 
It can be used to analyze how efficiently an algorithm has been parallelized.
}
of $A$ with respect to $B$.\footnote{To achieve a fair comparison of the algorithmic aspects of $A$ and $B$, the algorithmic speedup is often computed over their \emph{sequential} executions~\cite{hennessy2011computer}.
In view of today's ubiquitous parallelism, this perspective may need to be reconsidered, though.
}
To summarize multiple ratios, one can use the geometric mean~\cite{bixby2002solving}:
\[
  \mathrm{GM}(\text{speedup}) = \left(\prod_{i = 1}^{\text{\# of values}}{\text{speedup on instance }i}\right)^{\frac{1}{\text{\# of values}}}
\]
as it has the fundamental property that
$\mathrm{GM}(\text{speedup}) =
\frac{\mathrm{GM}(\text{running times of }A)}{\mathrm{GM}(\text{running times of }B)}$.
	Which mean is most appropriate for which aggregation is a matter of some discussion~\cite{mitchell200488,DBLP:journals/cacm/Smith88,DBLP:journals/cacm/FlemingW86}.

\subsubsection{Solution Quality}
        
Next, we discuss metrics for \emph{solution quality}. Here,
the correct measurements are naturally problem-specific;
often, there is no single quality indicator but multiple orthogonal indicators are used.

\paragraph*{Empirical vs. Worst-case Solution Quality}
As mentioned in the introduction, worst-case guarantees proven in theoretical models
are rarely approached in real-world settings.
For example, the accuracy of the ABRA algorithm for betweenness approximation
has been observed to be always below the
fixed absolute error, even in experiments where this was only guaranteed
for 90\% of all instances~\cite{riondato2018abra}.
Thus, experimental comparisons should include also metrics for which theoretical
guarantees are known.

\paragraph*{Comparing Against (Unknown) Ground Truth}
For many problems and instances beyond a certain size, \emph{ground truth}
(in the sense of the exact value of a centrality score or the true community structure of a network)
is neither known nor feasible to compute.
For betweenness centrality, however, AlGhamdi \etal~\cite{alghamdi2017} have computed exact scores
for large graphs by investing into considerable supercomputing time. 
The absence of ground truth or exact values, in turn, clearly requires the comparison to 
other algorithms in order to evaluate an algorithm's solution quality.

\section{Guidelines for the Experimental Pipeline}
\label{sec:exp-pipeline}

Organizing and running all the required experiments can be a time-consuming activity, especially
if not planned and carried out carefully. That is why we propose techniques and ideas 
to orchestrate this procedure efficiently.
The experimental pipeline can be divided into four phases.
In the first one, we finalize the algorithm's implementation as well as the scripts/code for the experiments themselves.\footnote{It is important to use scripts or some external tool in order to automate the experimental pipeline. 
This also helps to reduce human errors and  simplifies repeatability and replicability.
}
Next, we submit the experiments for execution (even if the experiments  are to be executed locally,
we advise to use some scheduling/batch system).
In the third phase, the experiments run and create the output files. Finally, we parse the output files 
to gather the information about the relevant metrics.

\subsection{Implementation Aspects}
\label{sub:implementation}
	Techniques for implementing algorithms are beyond the scope for
	this paper; however, we give an overview of tooling that should be
	used for developing algorithmic code.

	Source code should always be stored in version control systems (VCS);
  nowadays, the most commonly used VCS is Git~\cite{git}. For	scientific experiments, a VCS should also be used to version scripts that
	drive experiments, valuable raw experimental data and evaluation scripts.
	Storing instance names and algorithm parameters of experiments in VCS is
	beneficial, \eg when multiple iterations of experiments are done
	due to the AE cycle.

	The usual principles for software testing
	(\eg unit tests and assertions) should be applied to ensure that code behaves as expected. 
	This is particularly important in growing projects where seemingly local changes can affect other project parts with which the developer is not very familiar.
	It is often advantageous to open-source code.\footnote{We acknowledge that open-sourcing code is
		not always possible, \eg due to intellectual property or political reasons.}
		The \emph{Open Source Initiative} keeps a list\footnote{\url{https://opensource.org/licenses/alphabetical}.} of approved open source licenses.
		An important difference is whether they require users to publish derived products under the same license.
		If code is open-sourced, we suggest well-known platforms
		like Github~\cite{github}, Gitlab~\cite{gitlab}, or Bitbucket~\cite{bitbucket} to host it.
		An alternative is to use a VCS server within one's own organization, which reduces
		the dependence on commercial interests.
	In an academic context, a better accessibility can have the benefit of a higher scientific impact of the algorithms.
		For long-term archival storage, in turn, institutional repositories may be necessary.
	
	        Naturally, code should be well-structured and documented to encourage further scientific participation.
                Code documentation highly benefits from documentation generator
                tools such as Doxygen.\footnote{\url{http://www.doxygen.nl}.}
	\emph{Profiling} is usually used to find bottlenecks and optimize
	implementations, \eg using tools such as the \texttt{perf} profiler on Linux, Valgrind \cite{nethercote2007valgrind} or a commercial profiler such as VTune~\cite{reinders2005vtune}.

\subsection{Repeatability, Replicability and Reproducibility}
\label{sub:reproducibility}	
Terminology differs between venues; the Association of Computing Machinery defines \emph{repeatability} as obtaining the same results when the same team repeats the experiments,
\emph{replicability} for a different team but the same programs and \emph{reproducibility} for the case of a reimplementation by a different team. 
Our recommendations are mostly concerned with replicability.

In a perfect world scenario, the behavior of experiments is completely determined by
their code version, command line arguments and configuration files.
From that point of view, the ideal case for replicability,
which is increasingly demanded by conferences and journals\footnote{For example, see the \emph{Replicated
Computational Results Initiative} of the \emph{Journal on Experimental
Algorithms}, \url{http://jea.acm.org}.} in
experimental algorithms, looks like this:
A single executable program automatically downloads or generates the input
files, compiles the programs, runs the experiments and recreates the plots and
figures shown in the paper from the results.

Unfortunately, in reality some programs are non-deterministic
and give different outputs for the same settings.
If randomization is used, this problem is usually avoided
by fixing an initial seed of a pseudo-random number generator.
This seed is just another argument to the program and can be handled like all others.
However, parallel programs might still cause genuine
non-determinism in the output, \eg if the computation depends on the order in which a large search space is 
explored~\cite{hamadi2012parallel_sat_challenges,DBLP:conf/ipps/KimmigMS17} 
or on the order in which messages from other processors arrive.\footnote{As an example, some associative 
calculations are not associative when implemented with floating point numbers~\cite{Goldberg1991floating}. 
In such a case, the order of several, say, additions, matters.}
If these effects are of a magnitude that they affect the final result, these experiments need to be repeated sufficiently often to cover the distribution of outputs.
A replication would then aim at showing that its achieved results are, while not identical, equivalent in practice.
For a formal way to show such \emph{practical equivalence}, see Section~\ref{sec:stats:equivalence}.

Implementations often depend on libraries or certain compiler versions.
This might lead to complications in later replications when dependencies are no longer available or incompatible with modern systems.
Providing a virtual machine image or container addresses this problem.

\subsection{Running Experiments}
\label{sub:submission-collection}
	\newcommand{\sourceline}[1]{\begin{center}#1\end{center}}

\begin{figure}[tb]
\scriptsize
\begin{verbatim}
instances:
  konect:
    - 'advogato'
    - 'ego-twitter'
    - 'ego-facebook'
    - 'ego-gplus'
  # ... (more instances follow)
configurations:
  - name: kadabra-1t
    args: ['./run', '--threads=1', 'kadabra', '@INSTANCE@']
    output: stdout
  - name: kadabra-2t
    args: ['./run', '--threads=2', 'kadabra', '@INSTANCE@']
    output: stdout
  # ... (more configurations follow)
  - name: rk
    args: ['./run', 'rk', '@INSTANCE@']
    output: stdout
\end{verbatim}
\caption{\exptool configuration (\texttt{experiments.yml}) for \kad}
\label{fig:kadabra_exptool}
\end{figure}
	
	Running experiments means to take care of many details: Instances need to be
	generated or downloaded, jobs need to be submitted to batch
	systems\footnote{Note that the exact submission mechanism is beyond the scope
		of this paper, as it heavily depends on the batch system in question.
		Nevertheless, our guidelines and tooling suggestions can easily be adapted
		to all common batch system, such as Slurm (\url{https://slurm.schedmd.com/}) or PBS (\url{https://www.pbspro.org/}).
        }
	or executed locally, running jobs need to be monitored,
	crashes need to be detected, crashing jobs need to be restarted without restarting
	all experiments, \etc
	To avoid human errors, improve reproducibilty and accelerate those tasks,
	scripts and tooling should be employed.

	To help with these recurring tasks, we provide as a supplement to this paper \exptool,
	a command-line tool to automate the
	aforementioned tasks
	(among others).\footnote{\exptool can be found at \url{https://github.com/hu-macsy/simexpal}.}
	This tool allows the user to manage
	instances and experiments, launch jobs and monitor the status of those jobs.
	While our tool is not the only possible way to automate these tasks,
	we do hope that it improves over the state of writing custom scripts for
	each individual experiment.
	\exptool is configured using a simple YAML~\cite{yaml} file and only requires a minimal
	amount of user-supplied code. To illustrate this concept,
	we give an example configuration in Figure~\ref{fig:kadabra_exptool}.
	Here, \texttt{run} is the only piece of user-supplied code that needs to be written.
	\texttt{run} executes the algorithm and prints the output
	(e.g. running times) to \texttt{stdout}.
	Given such a configuration file, the graph instances can be downloaded using
	\texttt{\expcmd instances download}.
	After that is done, jobs can be started using the command
	\texttt{\expcmd experiments launch}. \exptool takes care of not
	launching experiments twice and never overwrites existing output files.
	\texttt{\expcmd experiments list}
	monitors the progress of all jobs. If a job crashes,
	\texttt{\expcmd experiments purge} can be used to remove
	the corresponding output files. The next \texttt{launch} command
	will rerun that particular job.

\subsection{Structuring Output Files}
\label{sub:benchmark-output}
	Output files typically store three kinds of data:
	(i) \emph{experimental results}, \eg running times
	and measures of solution quality,
	(ii) \emph{metadata} that completely specify the parameters and
	the environment of the run, so that the run can be replicated
	(see Section~\ref{sub:reproducibility}),
	and (iii) \emph{supplementary data}, \eg the solution to the input problem
	that can be used to understand the algorithm's behavior and to verify
	its correctness.
	Care must be taken to ensure that output files are suitable
	for long-term archival storage (which is mandated by good
	scientific practices~\cite{dfg:data_guidelines}).
	Furthermore, carefully designing output files helps to
	accelerate experiments by avoiding unnecessary runs that did
	not produce all relevant information (\eg if the focus of experiments
	changes after exploration).
	
	Choosing which experimental results to output is problem-specific but
	usually straightforward. For metadata, we recommend to include
	enough information to completely specify the executed algorithm version, its parameters and input instance, as well as the computing
	environment. This usually involves the following data:
	The \emph{VCS commit hash} of the implementation and compiler(s) as well as all
	libraries the implementation depends on\footnote{Experiments should never run
		uncommitted code. If there are any uncommitted changes,
		we suggest to print a comment to the output file to ensure that the
		experimental data in question
		does enter a paper.}
	name (or path) of the \emph{input instance},
	values of \emph{parameters}
	of the algorithm (including \emph{random seeds}\footnote{If the implementation's behavior
		can be controlled by a large number
		of parameters, it makes sense to print
		command line arguments
		as well as relevant environment variables and (excerpts from) configuration files
		to the output file.}),
	\emph{host name} of the machine and
	\emph{current date and time}.\footnote{Date and time help to identify
	the context of the experiments based on archived output data.}
	Implementations that depend on hardware details
	(\eg parallel algorithms or external-memory algorithms)
	want to log \emph{CPU, GPU and/or memory configurations},
	as well as \emph{versions of relevant libraries} and drivers.

	The relevance of different kinds of supplementary data is highly
	problem-dependent. Examples include (partial) solutions to the input problem,
	numbers of iterations that an algorithm performs, snapshots of
	the algorithm's state at key points during the execution
	or measurements of the time spent in different parts of the algorithm.
	Such supplementary data is useful
	to better understand an algorithm's behavior, to verify the correctness
	of the implementation or to increase confidence in the experimental
	results (\ie that the running times or solution qualities reported
	in a paper are actually correct).
	If solutions are stored, automated correctness checks can be used
	to find and debug problems or to demonstrate that no such problems exist.

	The \emph{output format} itself should be chosen to be both
	\emph{human readable} and \emph{machine parsable}.
	Human readability is particularly important for long-term archival
	storage, as parsers for proprietary binary formats (and the knowledge of how to use them)
	can be lost over time.
	Machine parsability enables automated extraction of experimental results;
	this is preferable over manual extraction, which is inefficient and error-prone.
	Thus, we recommend structured data formats like YAML (or JSON~\cite{json}).
	Those formats can be easily understood by humans; furthermore,
	there is a large variety of libraries to process them in any
	commonly used programming language.
	If plain text files are used, we suggest to ensure that
	experimental results can be extracted by simple regular expressions or similar tools.

\begin{figure}
\scriptsize
\begin{verbatim}
info:
  commit: fef6c5ca
  date: '2018-09-14T12:21:55.497368'
  host: erle
iterations: 12598
parameters:
  delta: 0.1
  epsilon: 0.015
  seed: 0
run_time: 1.6034371852874756
topk_nodes:
- 156
- 45
- 596
# ... (more nodes follow)
topk_scores:
- 0.0651690744562629
- 0.04643594221304969
- 0.0349261787585331
# ... (more scores follow)
\end{verbatim}
\caption{Output of \kad example in YAML format}
\label{fig:kadabra_outfile}
\end{figure}

\paragraph{\kad Example}
        Let us apply these guidelines to our example
	of the \kad algorithm using \exptool with the YAML file format.
	For each instance, we report \kad's running time,
	the values of the parameters $\epsilon$ and $\delta$,
	the random seed that was used for the run,
	the number $\tau$ of samples that the run required and the top-25 nodes of the resulting
	betweenness centrality ranking and their betweenness scores.
	The number 25 here is chosen arbitrarily, as a good balance between the amount of information
	that we store to verify the plausibility of the results and the amount of space
	consumed by the output. To fully identify the benchmark setting, we also report
	the hostname, our git commit hash, our random generator seed and the current date and time.
	Figure \ref{fig:kadabra_outfile} gives an example how the resulting output file looks like.

\subsection{Gathering Results}
\label{sub:gather-results}
\begin{figure}
\scriptsize
\begin{verbatim}
import simex
import yaml
cfg = simex.config_for_dir('.')
res = cfg.collect_successful_results(yaml.load)
for algo in 'kadabra', 'rk':
    rts = [r['run_time'] for r in res if r['algo'] == algo and r['threads'] == 1]
    print(algo, sum(rts)/len(rts))
\end{verbatim}
\caption{Script to collect \kad output}
\label{fig:exptool_collect}
\end{figure}
When the experiments are done, one has to verify that all runs were successful.
Then, the output data has to be parsed to extract the desired experimental
data for later evaluation. Again, we recommend the use of tools for parsing.
In particular, \exptool offers a Python package to collect output data.
Figure~\ref{fig:exptool_collect} depicts a complete script that computes
average running times for different algorithms on the same set of instances,
using only seven lines of Python code.
Note that \exptool takes care of reading the output files and checking that
all runs indeed succeeded (using the function \texttt{collect\_successful\_results()}).

In our example, we assume that the output files are formatted as YAML (thus we use the function
\texttt{yaml.load()} to parse them). In case a custom format is used (\eg when reading output
from a competitor where the output format cannot be controlled), the user has to supply
a small function to parse the output. Fortunately, this can usually be done using
regular expressions (\eg with Python's \texttt{regex} module).

Now would also a good time to aggregate data appropriately (unless this has been taken care of before, also see Section~\ref{sub:metrics}).

\section{Visualizing Results}
\label{sec:viz}
After the experiments have finished, the recorded data and results need to be explored and analyzed before they can finally be reported. 
Data visualization in the form of various different plots is a helpful tool both in the exploration phase and also in the final communication of the experimental results and the formal statistical analysis.
The amount of data collected during the experiments is typically too large to report in its entirety and hence meaningful and faithful data aggregations are 
needed.\footnote{For future reference and repeatability, it may make sense to include relevant raw data in tables
in the appendix. But raw data tables are rarely good for illustrating trends in the data.}
While descriptive summary statistics such as means, medians, quartiles, variances, standard deviations, or correlation coefficients, as well as results from statistical testing like p-values or confidence intervals  provide a well-founded summary of the data, they do so, by design, only at a very coarse level.
The same statistics can even originate from very different underlying data:
a famous example is Anscombe's quartet~\cite{anscombe73}, a collection of four sets of eleven points in $\mathbb R^2$ with very different point distributions, yet (almost) the same summary statistics such as mean, variance, or correlation.\footnote{For example plots of the four point sets see \url{https://commons.wikimedia.org/w/index.php?curid=9838454}.}
It is a striking example of how important it can be not to rely on summary statistics alone, but to visualize experimental data graphically. A more recent example is the datasaurus.\footnote{\url{https://www.autodeskresearch.com/publications/samestats}.}

Here we discuss a selection of \emph{plot types} for data visualization with focus on algorithm engineering experiments, together with guidelines when to use which type of plot depending on the properties of the data.
For a more comprehensive introduction to data visualization, we refer the reader to some of the in-depth literature on the topic~\cite{h-dvpi-18,s-pdfea-02,t-vdqi-01}.
Furthermore, there are many powerful libraries and tools for generating data plots, 
\eg R\footnote{\url{https://www.r-project.org}.} and the ggplot2 package\footnote{\url{https://ggplot2.tidyverse.org}.}, gnuplot\footnote{\url{http://www.gnuplot.info}.}, matplotlib.\footnote{\url{https://matplotlib.org}.} Also mathematical software such as MATLAB\footnote{\url{https://www.mathworks.com/products/matlab.html}.} (or even spreadsheet tools) can generate various types of plots from your data.
For more details about creating plots in one of these tools, we refer to the respective user manuals and various available tutorials.

\begin{figure}
\centering
\includegraphics[width=\textwidth]{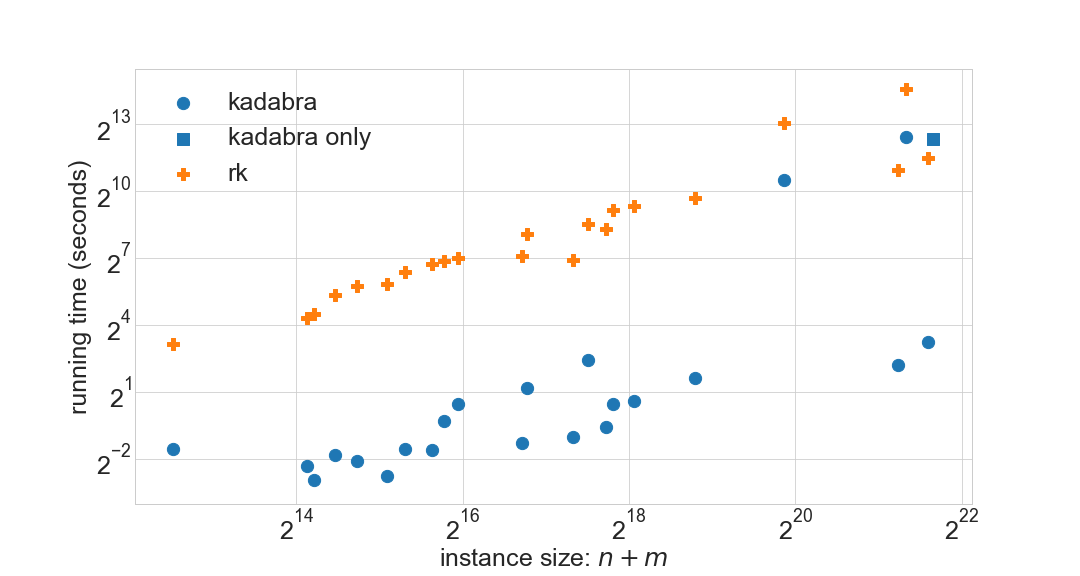}

\caption{Scatter plot for running times of \kad for $c=4375$ and \rk for all instances in the 
evaluation set. Both axes are in log scale.
Every point is the (arithmetic) mean running time for 5 repeated runs for each instance.
The one square represents the run of \kad for the network roadNet-TX for which \rk
did not finish within the cutoff time of 7 hours.
}
\label{fig:scatter}
\end{figure}

When presenting data in two-dimensional plots, \emph{marks} are the basic graphical elements or geometric primitives to show data. 
They can be points (zero-dimensional), lines (1D), or areas (2D). 
Each mark can be further refined and enriched by visual variables or \emph{channels} such as their position, size, shape, color, orientation, texture, \etc to encode different aspects of the data. 
The most important difference between those channels is that some are more suitable to represent categorical data (\eg graph properties, algorithms, data sources) by assigning different shapes or colors to them, whereas others are well suited for showing quantitative and ordered data (\eg input sizes, timings, quantitative quality measures) by mapping them to a spatial position, size, or lightness. 
For instance, in Figure~\ref{fig:scatter} we use blue circles as marks for one of the algorithms,
while for the other we use orange crosses.
Using different shapes makes sure that the plots are still readable if printed in grey-scale.
Not all channels are equally expressive and effective in encoding information for the human visual system, so that one has to carefully select which aspects of the data to map to which channels, possibly also using redundancy. 
For more details 
see the textbooks~\cite{m-vad-14} and~\cite{w-ivpd-12}.

\begin{wrapfigure}{R}{0.55\textwidth}
\centering
\includegraphics[width=0.55\textwidth]{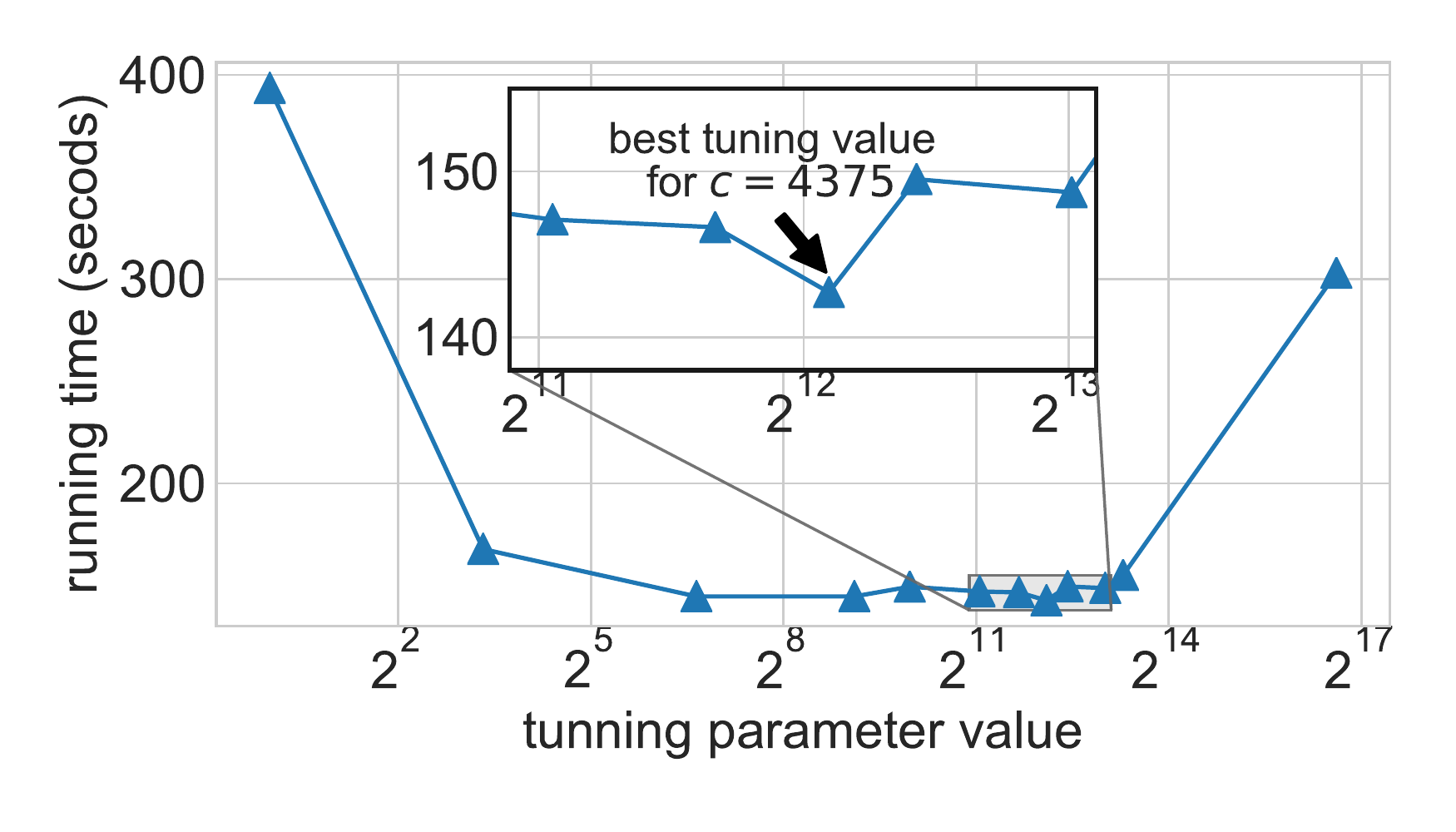}

\caption{ Tuning parameter search.
Every point is the (arithmetic) mean running time for all the instances in the tuning
set. Experiments on each instance are repeated 3 times for the respective value of the parameter.
The value that gives the best results is picked; here it is $4375$ as it provides the lowest running time of about $142$ seconds.
}
\label{fig:tunParam}
\end{wrapfigure}

As discussed in Section~\ref{sub:metrics}, the types of metrics from algorithmic experiments comprise two main aspects: running time data and solution quality data.
Both types of metrics can consist of absolute or relative values.
Typically, further attributes and parameters of the experimental data, of the algorithms, and of the input instances are relevant to include in a visualization to communicate the experimental findings. 
These can range from hardware-specific parameters, over the set of algorithms and possible algorithm-specific parameters, to instance-dependent parameters such as certain graph properties.
Depending on the experiment's focus, one needs to decide on the parameters to show in a plot and on how to map them to marks and channels.
Typically, the most important metric to answer the guiding research question (\eg running time or solution quality) is plotted along the y-axis.
The x-axis, in turn, is used for the most relevant parameter describing the instances (\eg instance size for a scalability evaluation or a graph parameter for evaluating its influence on the solution quality).
Additional parameters of interest can then be represented by using distinctly colored or shaped marks for different experimental conditions such as the respective algorithm, an algorithmic parameter or properties of the used hardware, see for example that, in Figures~\ref{fig:scatter} instance roadNet-TX
is plotted differently because \rk did not finish within the allocated time frame of 7 hours.

\begin{figure}
\includegraphics[width=\textwidth]{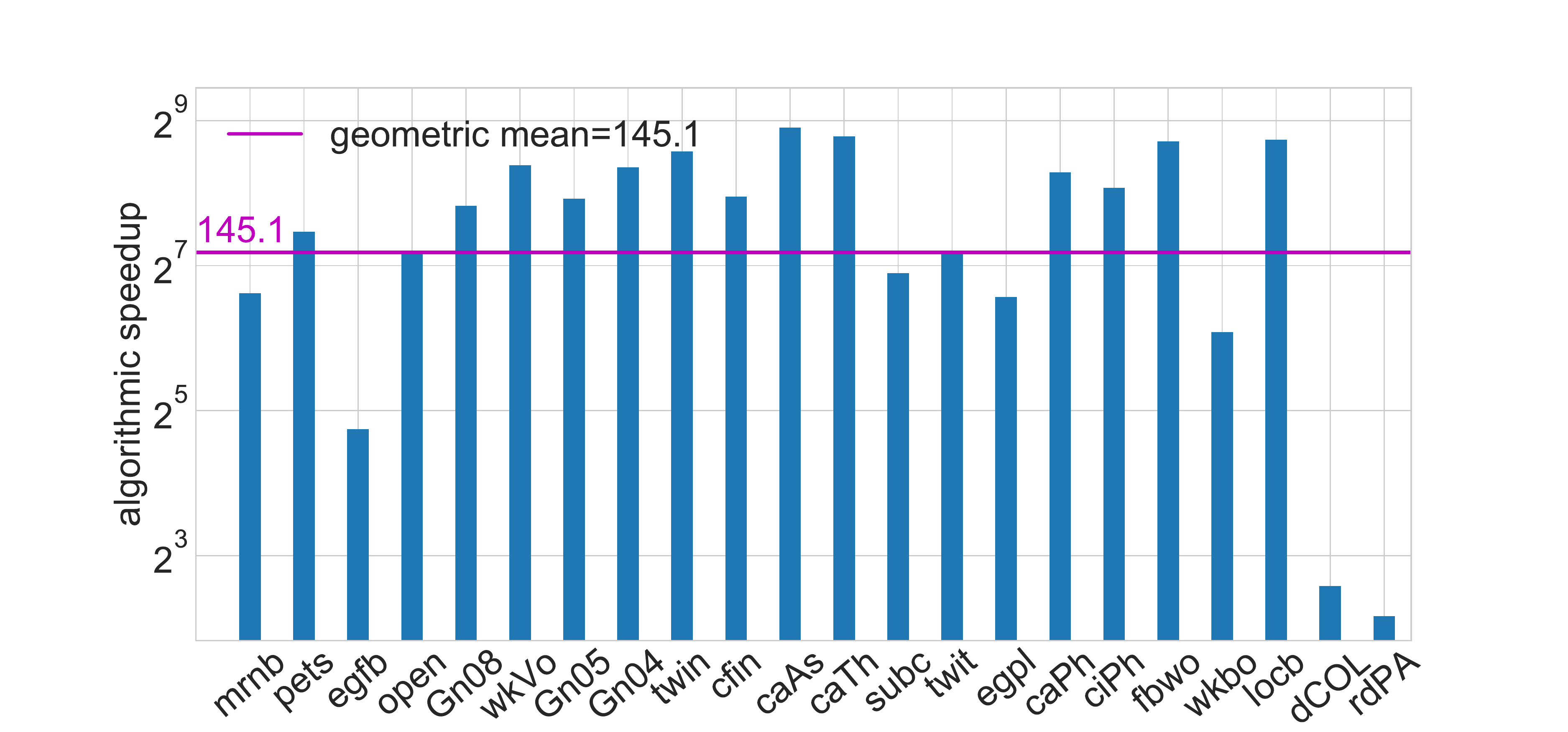}
\caption{The running time speedup of \kad over \rk for all the instances in the evaluation set
for $c=4375$. Every bar is the ratio of the arithmetic means for 5 repeated runs per instance.
The smallest speedup is $4.5$ for instance roadNet-PA (rdPA) 
and the largest speedup is $478.7$ for instance ca-AstroPh (caAs). 
The geometric mean for all speedups is around $145.1$.
Instances are sorted by their number of nodes. The y-axis is in log scale. } 
\label{fig:speedup}
\end{figure}

Before starting to plot the data, one needs to decide whether raw absolute data should be visualized (\eg running times or objective function values) or whether the data should be shown relative to some baseline value, or be normalized prior to plotting.
This decision typically depends on the specific algorithmic problem, the  experimental setup and the underlying research questions to be answered. 
For instance, when the experiment is about a new algorithm for a particular problem, the algorithmic speedup or possible improvement in solution quality with respect to an earlier algorithm may be of interest;
hence, running time ratios or quality ratios can be computed as a variable to plot -- as shown in Figure~\ref{fig:speedup} with respect to \kad and \rk.
Another possibility of data preprocessing is to normalize certain aspects of the experimental data before creating a plot.
For example, to understand effects caused by properties of the hardware, such as cache sizes and other memory effects, one may normalize running time measurements by the algorithm's time complexity in terms of $n$ and $m$, the number of vertices and edges of the input graph, and examine if the resulting computation times are constant or not.
A wide range of meaningful data processing and analysis can be done before showing the resulting data in a visualization.
While we just gave a few examples, the actual decision of what to show in a plot needs to be made by the designers of the experiment after carefully exploring all relevant aspects of the data from various perspectives. 
For the remainder of this section, we assume that all data values to be visualized have been selected.

A very fundamental plot is the \emph{scatter plot}, which maps two variables of the data (\eg size and running time)  onto the x- and y-axis, see Figure~\ref{fig:scatter}.
Every instance of the experiment produces its own point mark in the plot, by using its values of the two chosen variables as coordinates. 
Further variables of the data can be mapped to the remaining channels such as color, symbol shape, symbol size, or texture.
If the number of instances is not too large, a scatter plot can give an accurate visualization of the characteristics and trends of the data. 
However, for large numbers of instances, overplotting can quickly lead to scatter plots that become hard to read.

In such cases, the data needs to be aggregated before plotting, by grouping similar instances and showing summaries instead.
The simplest way to show aggregated data, such as repeated runs of the same instances, is to plot a single point mark for each group, \eg using the mean or median, and then optionally putting a vertical error bar on top of it showing one standard deviation or a particular confidence interval of the variable within each group.
Such a plot can be well suited for showing how running times scale with the instance size.
If the sample sizes in the experiment have been chosen to cover the domain well, one may amplify the salience of the trend in the data by linking the point marks by a line plot. 
However, this visually implies a linear interpolation between neighboring measurements and therefore should only be done if sufficiently many sample points are included and the plot is not misleading.
Obviously, if categorical data are represented on the x-axis, one should never connect the point marks by a line plot.

At the same time, the scale of the two coordinate axes is also of interest.
While a linear scale is the most natural and least confusing for human interpretation, some data sets contain values that may grow exponentially.
On a linear scale, this results in the large values dominating the entire plot and the small values disappear in a very narrow band.
In such cases, axes with a logarithmic scale can be used; however, this should always be made explicit in the axis labeling and the caption of the plot.

\begin{figure}[t]
\centering
\begin{subfigure}{.48\textwidth}
	\includegraphics[scale=0.21]{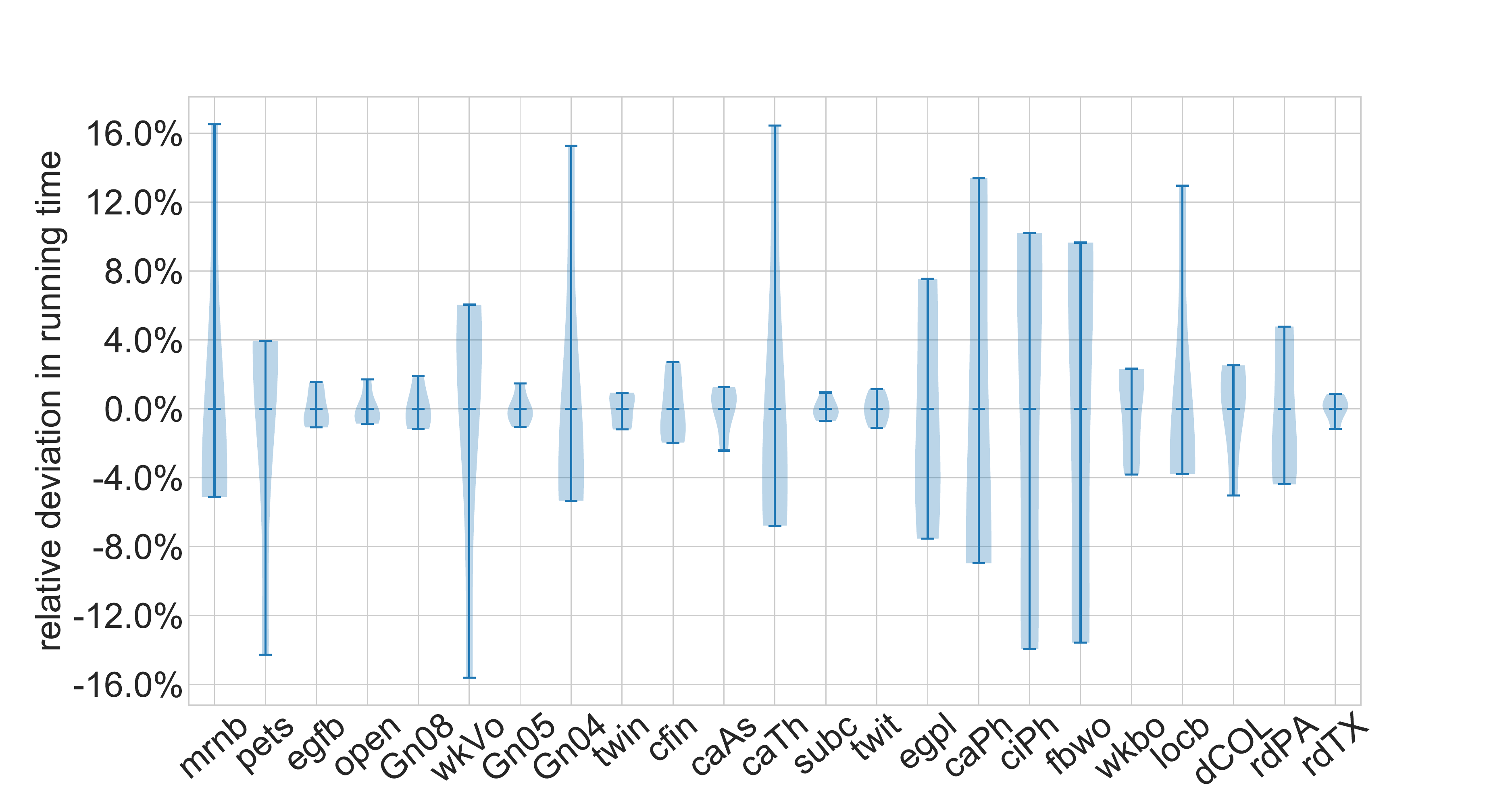}
	\caption{A violin plot for running time deviation of \kad. }
	\label{fig:box}
\end{subfigure}
\begin{subfigure}{0.48\textwidth}
	\includegraphics[scale=0.21]{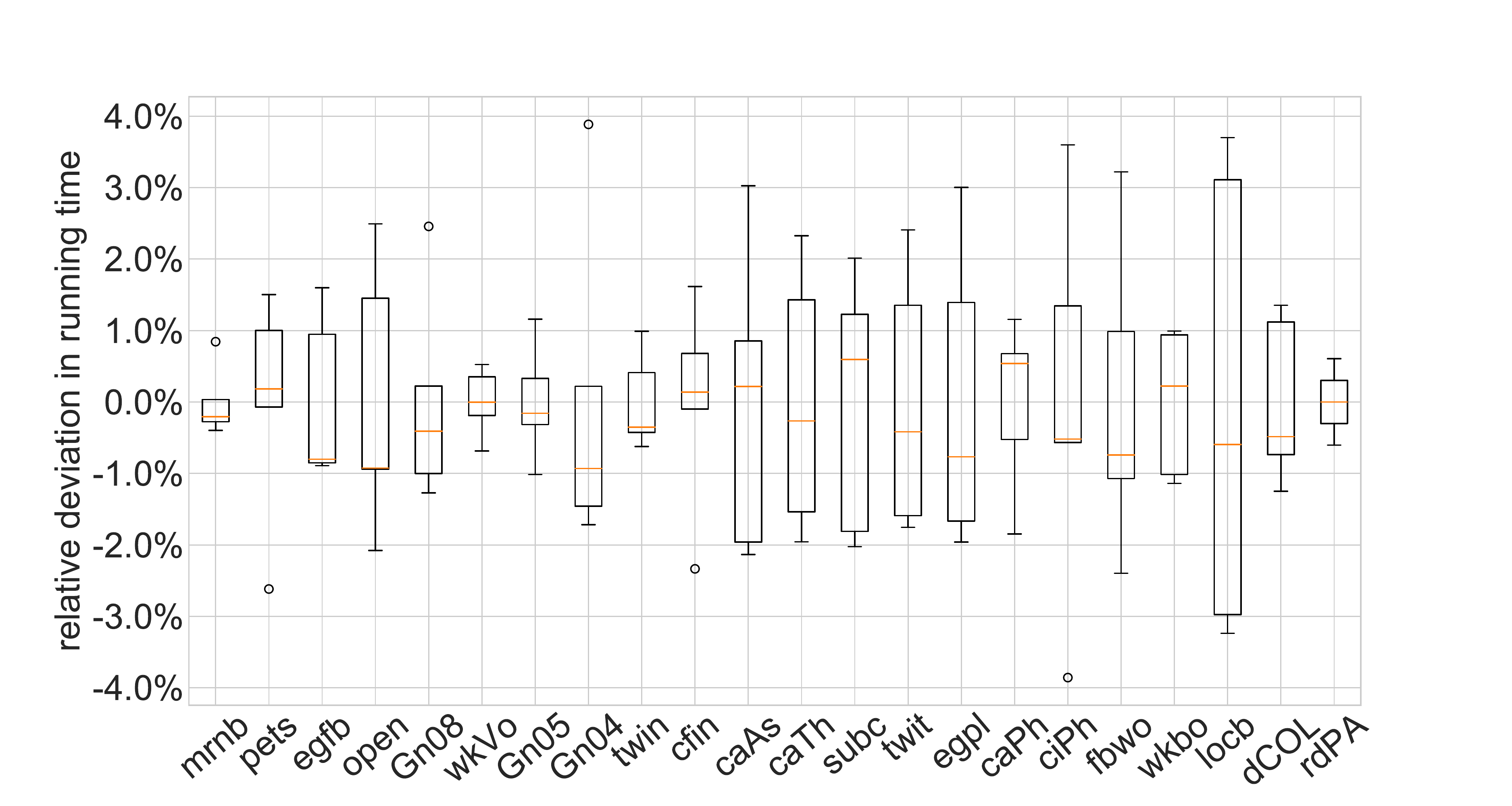}
	\caption{A box plot for running time deviation of \rk. The orange line indicates the median.}
	\label{fig:violin}
\end{subfigure}

\caption{  
Plots on variance of running times.
For each instance in the evaluation set, we perform 5 repeated runs and calculate the mean.
Shown in the plot are, for each run, the distance to the mean, divided by the mean.
This way we get the relative running time variance for every instance.
Here, we use a box and a violin plot for illustration purposes; in a scientific paper one of the two
should be chosen.
Notice also that instance roadNet-TX is missing for \rk since it did not finish within the 7 hours limit.
}
\end{figure}

A more advanced summary plot is the \emph{box plot} (or \emph{box-and-whiskers plot}), where all repeated runs of the same instance or sets of instances with equal or  similar size form a group, see Figure~\ref{fig:box}. 
This group is represented as a single glyph showing simultaneously the median, the quartiles, the minimum and maximum values or another percentile, as well as possibly outliers as individual marks.
If the x-axis shows an ordered variable such as instance size, one can still clearly see the scalability trend in the data as well as the variability within each homogeneous group of instances. 

\emph{Violin plots} take the idea of box plots even further and draw the contour of the density distribution of the variable mapped to the y-axis within each group, see Figure~\ref{fig:violin}.
It is thus visually more informative than a simple box plot, but also more complex and thus possibly more difficult to read.
When deciding for a particular type of plot, one has to explore the properties of the data and choose a plot type that is neither oversimplifying them nor more complex than needed.

Data from algorithmic experiments may also often be aggregated by some attribute into groups of different sizes. 
In order to show the distribution of the instances into these groups, bar charts (for categorical data) or histograms (for continuous data) can be used to visualize the cardinality of each group.
Such diagrams are often available in public repositories like KONECT.\footnote{For example, one can
find this information for moreno\_blogs in \url{http://konect.uni-koblenz.de/networks/moreno_blogs}.}
For a complex network, for example, one may want to plot the degree distribution of its nodes with the degree (or bins with a specific degree range) on the x-axis and the number of vertices with a particular degree (or in a particular degree range) on the y-axis. 
Such a histogram can then quickly reveal how skewed the degree distribution is.
Similarly, histograms can be useful to show solution quality ratios obtained by one or more algorithms by defining bins based on selected ranges of quality ratios.
Such a plot quickly indicates to the reader what percentage of instances could be solved within a required quality range, \eg  at least $95\%$ of the optimum. 
of the algorithm(s) and the $y$-axis displays the quality. This allows 

A single plot can contain multiple experimental conditions simultaneously.
For instance, when showing running time behavior and scalability of a new algorithm compared to previous approaches, a single plot with multiple trend lines in different colors or textures or with visually distinct mark shapes can be very useful to make comparisons among the competing algorithms.
Clearly, a legend needs to specify the mapping between the data and the marks and channels in the plot.
Here it is strongly advisable to use the same mapping if multiple  plots are used that all belong together.
But, as a final remark, bear in mind the size and resolution of the created figures. 
Avoid clutter and ensure that your conclusion remains clearly visible!

\section{Evaluating Results with Statistical Analysis}
\label{sec:statistical-analysis}
Even if a result looks obvious\footnote{It fulfills the ``inter-ocular trauma test'', as the saying goes, the evidence hitting you between the eyes.}, it can benefit from a statistical analysis to quantify it, especially if random components or heuristics are involved.
The most common questions for a statistical analysis are:
\begin{itemize}
\item Do the experimental results support a given hypothesis, or is the measured difference possibly just random noise? This question is addressed by \emph{hypothesis testing}.
\item How large is a measured effect, \ie which underlying real differences are plausible given the experimental measurements? This calls for \emph{parameter estimation}.
\item If we want to answer the first two questions satisfactorily, how many data points do we need? This is commonly called \emph{power analysis}.
As this issue affects the planning of experiments, we discussed it already in Section~\ref{sub:variance}.
\end{itemize}

The two types of hypotheses we discussed in Section~\ref{sub:guidelines:hypotheses} roughly relate to hypothesis tests and parameter estimation.
Papers proposing new algorithms mostly have hypotheses of the first type: The newly proposed algorithm is faster, yields solutions with better quality or is otherwise advantageous.

In many empirical sciences, the predominant notion has been \emph{null hypothesis significance testing} (NHST),
in which a statistic of the measured data is compared with the distribution of this statistic under the \emph{null hypothesis}, an assumed scenario where all measured differences are due to chance.
Previous works on statistical analysis of experimental algorithms, including the excellent overviews of McGeoch~\cite{McGeoch12} and Coffin and Saltzmann~\cite{Coffin00}, use this paradigm.

Due to some limitations of the NHST model, a shift towards \emph{parameter estimation}~\cite{anderson2000null,doi:10.1080/01973533.2015.1012991,wasserstein2016asa,lash2017harm} and also Bayesian methods~\cite{doi:10.1177/0956797613504966,kruschke2018bayesian,murtaugh2014defense} is taking place in the statistical community.

We aim at applying the current state of the art in statistical analysis to algorithm engineering, but since no firm consensus has been reached~\cite{wasserstein2016asa}, we discuss both frequentist and Bayesian approaches.
As an example for null hypothesis testing, we investigate whether the \kad and the \rk algorithms give equivalent results for the same input graphs.
While this \emph{equivalence test} could also be performed using Bayesian inference, we use a null hypothesis test for illustration purposes.

It is easy to see from running time plots (\eg Figure~\ref{fig:scatter}) that \kad is faster than \rk.
To quantify this speedup, we use Bayesian methods to infer plausible values for the algorithmic speedup and different scaling behavior.
We further evaluate the influence of the graph diameter on the running time.

\subsection{Statistical Model}
\label{sec:stats:model}
\newcommand{\prob}[1]{\ensuremath{{P}(#1)}}
\newcommand{\likelihood}[2]{\ensuremath{\mathcal{L}(#1|#2)}}

A statistical model defines a family of probability distribution over experimental measurements.
Many experimental running times have some degree of randomness:
Caching effects, network traffic and influence of other processes contribute to this, sometimes the tested algorithm is randomized itself.
Even a deterministic implementation on deterministic hardware can only be tested on a finite set of input data, representing a random draw from the infinite set of possible inputs.
A model encodes our \emph{assumptions} about how these sources of randomness combine to yield the distribution of outputs we see.
If, for example, the measurement error consists of many additive and independent parts, the central limit theorem justifies a normal distribution.

To enable useful inferences, the model should have at least one free \emph{parameter} corresponding to a quantity of interest.
Any model is necessarily an abstraction, as expressed in the aphorism ``all models are wrong, but some are useful''~\cite{Box1976models}.

\subsubsection{Example}
Suppose we want to investigate whether \kad scales better than \rk on inputs of increasing size.
Figure~\ref{fig:scatter} shows running times of \kad and \rk with respect to the instance size.
These are spread out, implying either that the running time is highly variable, or that it depends on aspects other than the instance size.

A companion Jupyter notebook including this example and the following inferences is included in the supplementary materials.

In general, running times are modeled as functions of the input size, sometimes with additional properties of the input.
The running time of \kad, for example, possibly depends on the diameter of the input graph.
Algorithms in network analysis often have polynomial running times where a reasonable upper bound for the leading exponent can be found.
Thus, the running time can be modeled as such a polynomial, written as $a + bn + cn^2 + dn^3 + \ldots + \alpha \log n + \beta \log \log n + \ldots $, with the unknown coefficients of the polynomial being the free parameters of the model.

However, a large number of free model parameters makes inference difficult.
This includes the danger of overfitting, \ie inferring parameter values that precisely fit the measured results but are unlikely to generalize.

To evaluate the scaling behavior, it is thus often more useful to focus on the largest exponent instead.
Let $T_A(n)$ be the running time of the implementation of \rk and $T_B(n)$ the running time of the implementation of \kad on inputs of size $n$,
with unknown parameters $\alpha_A, \alpha_B, \beta_A$ and $\beta_B$.\footnote{For estimating asymptotic upper bounds, see the work of McGeoch \etal~\cite{McGeochP00} on curve bounding.}
\begin{align*}
 T_A(n) = \alpha_A \cdot n^{\beta_A} \cdot \epsilon\\
 T_B(n) = \alpha_B \cdot n^{\beta_B} \cdot \epsilon
\end{align*}
The term $\epsilon$ explicitly models the error; it can be due to variability in inputs (some instances might be harder to process than others of the same size) and measurement noise (some runs suffer from interference or caching effects).
Since harder inputs are not constrained to additive difficulty and longer runs have more opportunity to experience adverse hardware effects, 
we choose a multiplicative error term.\footnote{Summands with smaller exponents are also subsumed within the error term. If they have a large effect, an additive error might reflect this more accurately.}
Taking the logarithms of both sides makes the equations linear:
\begin{align}
 \log(T_A(n)) = \log(\alpha_A) + \beta_A \log(n) + \log(\epsilon)\label{eq:log-transformed-a} \\
 \log(T_B(n)) = \log(\alpha_B) + \beta_B \log(n) + \log(\epsilon)\label{eq:log-transformed-b}
\end{align}

A commonly chosen distribution for additive errors is Gaussian, justified by the central limit theorem~\cite{polya1920zentralen}.
Since longer runs have more exposure to possibly adverse hardware or network effects we consider multiplicative error terms to be more likely and use a log-normal distribution.\footnote{In some cases, it might even make sense to use a hierarchical model with two different error terms: One for the input instances, the other one for the differences on the same input.}
Since the logarithm of the log-normal distribution is a normal (Gaussian) distribution, Equations~\ref{eq:log-transformed-a} and \ref{eq:log-transformed-b} can be rewritten as normally distributed random variables:
\begin{align}
 \log(T_A(n)) \sim \mathcal{N}(\log(\alpha_A) + \beta_A  \log(n),\ \sigma_\epsilon^2)\label{eq:log-normal-a} \\
 \log(T_B(n)) \sim \mathcal{N}(\log(\alpha_B) + \beta_B  \log(n),\ \sigma_\epsilon^2)
\end{align}

This form shows more intuitively the idea that a statistical model is \emph{a set of probability measures on experimental outcomes}.
In this example, the set of probability measures modeling the performance of an algorithm are parametrized by the tuple $(\alpha, \beta, \sigma) \in \mathbb{R}^3$.

A problem remains if the input instances are very inhomogeneous.
In that case, both $A$ and $B$ have a large estimated variance ($\sigma^2$).
Even if the variability of performance \emph{on the same instance} is small, any genuine performance difference might be wrongly attributed to the large \emph{inter-instance} variance.
This issue can be addressed with a combined model as recommended by Coffin and Saltzmann~\cite{Coffin00}, in which the running time $T_A(x)$ of $A$ is a function of the running time $T_B(x)$ of $B$ \emph{on the same instance} $x$:
\begin{equation}
 \log(T_A(x)) \sim \mathcal{N}(\log(\alpha_{A/B}) + \beta_{A/B}  \log(T_B(x)),\ \sigma^2)\label{eq:relative-time-model}
\end{equation}
For a more extensive overview of modeling experimental algorithms, see \cite{McGeoch12}.

\subsubsection{Model Fit}
Several methods exist to infer parameters when fitting a statistical model to experimental data.
The most well-known is arguably the maximum-likelihood fit, choosing parameters for a distribution that give the highest probability for the observed measurements.
In the case of a linear model with a normally distributed error, this is equivalent to a least-squares fit~\cite{doi:10.1080/01621459.1976.10481508}.
Such a fit yields a single estimate for plausible parameter values.

Given random variations in the measurement, we are also interested in the \emph{range} of plausible values, quantifying the uncertainty involved in measurement and instance selection.
This is covered in Sections \ref{sec:stats:confidence-intervals} and \ref{sec:stats:bayesian-inference}.

\subsection{Formalizing Hypotheses}
\label{ss:stats:hypotheses}
Previously (Section~\ref{sub:guidelines:hypotheses}), we discussed two types of hypotheses.
The first type is that a new algorithm is better in some aspect than the state of the art.
The second type claims insight into how the behavior of an algorithm depends on settings and properties of the input.
We now express these same hypotheses more formally, as statements about the parameters in a statistical model.
The two types are then related to the statistical approaches of \emph{hypothesis testing} and \emph{parameter estimation}.

Considering the scaling model presented in Equation~(\ref{eq:relative-time-model}) and the question whether implementation A scales better than implementation B,
the parameter in question is the exponent $\beta_{A/B}$.
The hypothesis that \emph{A} scales better than \emph{B} is equivalent to $\log(\beta_{A/B}) < 0$; both scaling the same implies $\log(\beta_{A/B}) = 0$.
Note that the first hypothesis does not imply a fully specified probability distribution on $\beta_{A/B}$, merely restricting it to the negative half-plane.
The hypothesis of $\log(\beta_{A/B}) = 0$ does completely specify such a distribution (\ie a point mass of probability 1 at 0), which is useful for later statistical inference.

\subsection{Frequentist Methods}
\label{sec:stats:frequentist}
Frequentist statistics defines the probability of an experimental outcome as the \emph{limit of its relative frequency} 
when the number of experiments trends towards infinity.
This is usually denoted as the classical approach.

\subsubsection{Null Hypothesis Significance Testing}
\label{sec:stats:nhst}
As discussed above, \emph{null hypothesis significance testing} evaluates a proposed hypothesis by contrasting it with a \emph{null hypothesis}, which states that no true difference exists and the observed difference is due to chance.
As an example application, we compare the approximation quality of \kad and \rk.
From theory, we would expect higher approximation errors from \kad, since it samples fewer paths.
We investigate whether the measured empirical difference supports this theory, or could also be explained with random measurement noise, \ie with the null hypothesis (denoted with $H_0$) of both algorithms having the same distribution of errors and just measuring higher errors from \kad by coincidence.
Here it is an advantage that the proposed alternate hypothesis (\ie the distributions are meaningfully different) does not need an explicit modeling of the output distribution, as the distribution of differences in our case does not follow an easily parameterizable distribution.

When deciding whether to reject a null hypothesis (and by implication, support the alternate hypothesis), it is possible to make one of two errors: (i) rejecting a null hypothesis, even though it is true (false positive),
(ii) failing to reject a false null hypothesis (false negative).
In such probabilistic decisions, the error rate deemed acceptable often depends on the associated costs for each type of error.
For scientific research, Fisher suggested that a false positive rate of 5\% is acceptable~\cite{fisher1992statistical}, and most fields follow that suggestion.
This threshold is commonly denoted as $\alpha$.

Controlling for the first kind of error, the $p$-value is defined as the probability that a \emph{summary statistic} as extreme as the observed one would have occurred given the null hypothesis~\cite{young2005essentials}.
Please note that this is \emph{not} the probability $P(H_0 | \text{observations})$, \ie the probability that the null hypothesis is true given the observations.

For practical purposes, a wide range of \emph{statistical hypothesis tests} have been developed, which aggregate measurements to a summary statistic and often require certain conditions.
For an overview of which of them are applicable in which situation, see the excellent textbook of Young and Smith~\cite{young2005essentials}.
In our example the paired results are of very different instances and clearly not normally distributed.
We thus avoid the common t-test and use a Wilcoxon test of pairs~\cite{wilcoxon1945individual} from the SciPy~\cite{scipy} stats module, yielding a $p$-value of $5.9~\cdot 10^{-5}$, see cell 14 of the statistics notebook in the supplementary materials.
Since this is smaller than our threshold $\alpha$ of 0.05, one would thus say that this result allows us to reject the null hypothesis at the level of $5.9~\cdot 10^{-5}$.
Such a difference is commonly called \emph{statistically significant}.
To decide whether it is actually significant in practice, we look at the magnitude of the difference:
The error of \kad is about one order of magnitude higher for most instances, which we would call significant.

\paragraph{Multiple Comparisons}
The NHST approach guarantees that of all false hypotheses, the expected fraction that seem significant when tested is at most $\alpha$.
Often though, a publication tests more than one hypotheses.
For methods to address this problem and adjust the probability that \emph{any} null hypothesis in an experiment is falsely rejected (also called the \emph{familywise error rate}), see Bonferroni~\cite{dunn1961multiple} and Holm~\cite{holm1979bonferroni}.

\subsubsection{Confidence Intervals}
\label{sec:stats:confidence-intervals}
One of the main criticism of NHST is that it ignores effect sizes;
the magnitude of the $p$-value says nothing about the magnitude of the effect.
More formally, \emph{for every true effect with size $\epsilon > 0$ and every significance level $\alpha$,
there exists an $n_0$ so that all experiments containing $n \geq n_0$ measurements are likely to reject the null hypothesis at level $\alpha$.}
Following this, Coffin and Saltzmann~\cite{Coffin00} caution against overly large data sets - a recommendation which comes with its own set of problems.

The statistical response to the problem of small but meaningless $p$-values with large data sets is a shift away from hypothesis testing to \emph{parameter estimation}.
Instead of asking whether the difference between populations is over a threshold, the difference is quantified as a parameter in a statistical model, see also Section~\ref{ss:stats:hypotheses}.
As Kruschke \etal~\cite{kruschke2018bayesian} put it, the null hypothesis test asks whether the null value of a parameter would be rejected at a given significance level.
A \emph{confidence interval} merely asks which other parameter values would not be rejected~\cite{kruschke2018bayesian}.
We refer to ~\cite{smithsonCI,neyman1937x} for a formal definition and usage guidelines.

\subsection{Bayesian Inference}
\label{sec:stats:bayesian-inference}
Bayesian statistics defines the probability of an experimental outcome as the \emph{uncertainty of knowledge about it}.
Bayes's theorem gives a formal way to update probabilities on new observations.
In its simplest form for discrete events and hypotheses, it can be given as:
\begin{equation}
 \prob{H_1 | \text{A}} = \frac{\prob{\text{A} | H_1}\prob{H_1}}{ \prob{\text{A}}}
\end{equation}
When observing outcome $A$, the probability of hypothesis $H_1$ is proportional to the probability of $A$ conditioned on $H_1$ multiplied by the \emph{prior probability} $\prob{H_1}$.
The conditional probability $\prob{A | H_1}$ of an outcome $A$ given an hypothesis $H_1$ is also called the \emph{likelihood} of $H_1$.
The prior probability reflects the estimation before making observations, based on background knowledge.

Extended to continuous distributions, Bayes's rule allows to combine a statistical model with a set of measurements and a \emph{prior probability distribution} over parameters to yield a \emph{posterior probability distribution} over parameters.
This posterior distribution reflects both the uncertainty in measurements and possible prior knowledge about parameters.
A thorough treatment is given by Gelman \etal~\cite{gelman2013bayesian}.

For our example model introduced in Section~\ref{sec:stats:model}, we model the running times of implementation B as a function of the time of implementation A, as done in Equation~(\ref{eq:relative-time-model}):
\[
\log(T_A(n)) \sim \mathcal{N}(\alpha_{A/B} + \beta_{A/B} \cdot \log(T_B(n)), \sigma_{A/B}^2)
\]
This defines the likelihood function as a Gaussian noise with variance $\sigma^2$.
Since this variance is unknown, we keep it as a model parameter.
As we have no specific prior information about plausible values of $\alpha$ and $\beta$, we define the following vague prior distributions:
\begin{align*}
 \alpha_{A/B} \sim \mathcal{N}(0,10),\  
 \beta_{A/B} \sim \mathcal{N}(1,10),\ 
 \sigma_{A/B} \sim \mathrm{Inv\-Gamma}(1,1)
\end{align*}
The first two distributions represent our initial -- conservative -- belief that the two implementations are equivalent in terms of scaling behavior and constants.
We model the variance of the observation noise $\sigma^2$ as an inverse gamma distribution instead of a normal distribution, since a variance cannot be negative.

Figure~\ref{fig:bayes-modeling} shows a listing to compute and show the posterior distribution of these three parameters using SciPy~\cite{scipy} and PyMC3~\cite{salvatier2016probabilistic}.
\begin{verbbox}[\scriptsize]
import pymc3 as pm
from scipy import optimize

basic_model = pm.Model()

with basic_model:
	alpha = pm.Normal('alpha', mu=0, sd=10)
    beta = pm.Normal('beta', mu=0, sd=10)
    sigma = pm.InverseGamma('sigma', alpha=1,beta=1)

    mu = alpha + beta*logTimeRK
    
    Y_obs = pm.Normal('Y_obs', mu=mu, sd=sigma, observed=logTimeKadabra)

with basic_model:
    # approximate posterior distribution with 10000 samples
    trace = pm.sample(10000)

pm.summary(trace)
\end{verbbox}
\begin{figure}
\centering
\theverbbox
\caption{Example listing for inference of parameters $\alpha_{A/B}$, $\beta_{A/B}$ and $\sigma_{A/B}$.
}
\label{fig:bayes-modeling}
\end{figure}
\begin{table}
\centering
\begin{tabular}{llll}
 & HPD 2.5 & Mean & HPD 97.5\\
\toprule
$\alpha_{A/B}$ & -6.87 & -5.22 & -3.58\\
$\beta_{A/B}$ & 0.70 & 1.01 & 1.29\\
$\sigma_{A/B}$ & 0.80 & 1.13 & 1.54\\
\bottomrule
\end{tabular}
\caption{Posterior distribution of parameter values.
}
\label{table:bayesian-posterior-results}
\end{table}
Results are listed in Table~\ref{table:bayesian-posterior-results}.
The interval of \emph{Highest Probability Density} (HPD) is constructed to contain 95\% of the probability mass of the respective posterior distribution.
It is the Bayesian equivalent of the confidence interval (Section~\ref{sec:stats:confidence-intervals}) and also called \emph{credible interval}.
The most probable values for $\alpha_{A/B}$ and $\beta_{A/B}$ are -5.22 and 1.01, respectively.
Taking measurement uncertainty into account, the true values are within the intervals $[-6.87, -3.58]$ respective $[0.70, 1.29]$ with 95\% probability.
This shows that \kad is faster on average, but results about the relative scaling behavior are inconclusive.
While the average for $\beta_{A/B}$ is 1.01 and suggests similar scaling, the interval $[0.7, 1.29]$ neither excludes the hypothesis that \kad scales better, nor the hypothesis that it scales worse.

\subsubsection{Equivalence Testing}
\label{sec:stats:equivalence}
Computing the highest density interval can also be used for hypothesis testing.
In Section~\ref{sec:stats:nhst} we discussed how to show that two distributions are different.
Sometimes, though, we are interested in showing that they are sufficiently \emph{similar}.
An example would be wanting to show that two sampling algorithms give the same distribution of results.
This is not easily possible within the NHST paradigm, as the two answers of a classical hypothesis test are ``probably different'' and ``not sure''.

This problem can be solved by calculating the posterior distribution of the parameter of interest and defining a \emph{region of practical equivalence} (ROPE), which covers all parameter values that are effectively indistinguishable from 0.
If $x\%$ of the posterior probability mass are in the region of practical equivalence, the inferred parameter is practically indistinguishable with probability $x\%$.
If $x\%$ of the probability mass are outside the ROPE, the parameter is meaningfully different with probability $x\%$.
If the intervals overlap, the observed data is insufficient to come to either conclusion.
In our example, the scaling behavior of two algorithms is equivalent if the inferred exponent modeling their relative running times is more or less 1.
We could define practical equivalence as $\pm 5\%$, resulting in a region of practical equivalence of $[0.95, 1.05]$.
The interval $[0.7, 1.29]$ containing $95\%$ of the probability mass for $\beta$ is neither completely inside $[0.95, 1.05]$ nor completely outside it, implying that more experiments are needed to come to a conclusion.

\subsubsection{Bayes Factor}
\label{sec:stats:bayes-factor}
\emph{Bayes factors} are a way to compare the \emph{relative fit} of several models and hypotheses to a given set of observations.
While NHST (Section~\ref{sec:stats:nhst}) evaluates the fit of observations to the null hypothesis and confidence intervals and credible intervals infer the range of plausible values of parameters in a model, the Bayes factor between two hypotheses gives the ratio between their posterior probabilities.
This \emph{probability ratio} of hypotheses is then:
\[
 \frac{\prob{H_1 | \text{obs}}}{\prob{H_0 | \text{obs}}}  = \frac{\prob{\text{obs} | H_1}}{ \prob{\text{obs} | H_0}} \ \frac{\prob{H_1}}{\prob{H_0}}
\]
Crucially, the ratio of prior probabilities, which is subjective, is a separate factor from the ratio of likelihoods, which is objective.
This objective part, the ratio $\frac{\prob{\text{obs} | H_1}}{ \prob{\text{obs} | H_0}}$, is called the \emph{Bayes factor}.

The first obvious difference to NHST is that calculating a Bayes Factor consists of comparing the fit of \emph{both} hypotheses to the data, not only the null hypothesis.
It thus requires that an alternate hypothesis is stated explicitly, including a probability distribution over observable outcomes.
If the alternative hypothesis is meant to be vague, for example just that two distributions are different, an uninformative prior with a high variance should be used.
However, specific hypotheses like ``the new algorithm is at least 20\% faster'' can also be modeled explicitly.

This explicit modeling allows inference in both directions;
using NHST, on the other hand, one can only ever infer that a null hypothesis is unlikely or that the data is insufficient to infer this.
Using Bayes factors, it is possible to infer that $H_1$ is more probable than $H_0$,
or that the observations are insufficient to support this statement, or that $H_0$ is more probable than $H_1$.

In the previous running time analysis, we hypothesized a better scaling behavior, which was not confirmed by the experimental measurements.
However, the graphs with high diameter are larger than average and a cursory complexity analysis of the \kad algorithm suggests that the diameter has an influence on the running time.
Might the relative scaling of \kad and \rk depend on the running time?

To answer this question, we compare the fit of two models:
The first model is the same as discussed earlier (Equation~\ref{eq:relative-time-model}),
it models the expected running time of \kad on instance $x$ as $e^{\alpha} \cdot T_{\mathrm{RK}}(x)^{\beta}$, where $T_{\mathrm{RK}}(x)$ is the running time of \rk on the same instance.
The second model has the additional free parameter $\gamma$, controlling the interaction between the diameter and running times:
\begin{equation}
\log(T(x)) \sim \mathcal{N}(\log(\alpha) + \beta \log(T_{\mathrm{RK}}(x)) + \gamma \log(\text{diam}_x), \,\ \sigma_\epsilon).
\label{eq:log-transformed-diameter}
\end{equation}

Comparing for example the errors of a least-squares fit of the two models would not give much insight, since including an additional free parameter in a model almost always results in a better fit.
This does not have to mean that the new parameter captures something interesting. 

Instead, we calculate the Bayes factor, for which we integrate over the prior distribution in each model.
Since this integral over the prior distribution also includes values of the new parameter which are not a good fit, models with too many additional parameters are automatically penalized.
Our two models are similar, thus we can phrase them as a hierarchical model with the additional parameter controlled by a boolean random variable, see Listing~\ref{listing:bayes-factor}.

\begin{verbbox}[\scriptsize]
import pymc3 as pm
from scipy import optimize

basic_model = pm.Model()

with basic_model:
    
    pi = (0.5, 0.5)
    
    selected_model = pm.Bernoulli('selected_model', p=pi[1])

    alpha = pm.Normal('alpha', mu=0, sd=10)
    beta = pm.Normal('beta', mu=0, sd=10)
    gamma = pm.Normal('gamma', mu=0, sd=10)
    sigma = pm.InverseGamma('sigma', alpha=1, beta=1)

    mu = alpha + beta*logTimesRK + gamma*logDiameters*selected_model
    
    Y_obs = pm.Normal('Y_obs', mu=mu, sd=sigma, observed=logTimeKadabra)

\end{verbbox}

\begin{figure}
\centering
\theverbbox
\caption{Listing to compute the Bayes factor, the relative likelihood of two models, between the model including the graph diameter and the model with just the graph sizes.
The variable \emph{selected\_model} is an indicator for which model is selected.
}
\label{listing:bayes-factor}
\end{figure}

The posterior for the indicator variable \texttt{selected\_model} is $0.945$, yielding a Bayes factor of $\approx 17$ in support of including the diameter in the model.
We can thus conclude that it is very probable that the diameter has an influence on the relative scaling between \kad and \rk.
The inferred mean of the variable $\gamma$ is 0.79, meaning that higher diameter values lead to higher running times.

\subsection{Recommendations}
\label{sec:stats:recommendations}
Which statistical method is best, depends on what needs to be shown.
For almost all objectives, both Bayesian and frequentist methods exist, see Table~\ref{table:stats-methods}.
In experimental algorithmics, most hypotheses can be expressed as statements constraining parameters in a statistical model, \ie ``the average speedup of A over B is at least 20\%''.
Thus, in contrast to earlier statistical practice, we recommend to approach evaluation of hypotheses by parameter estimation and to only use the classical hypothesis tests when parameter estimation is not possible.
The additional information gained by parameter estimates has a couple of advantages.
For example, when using only hypothesis tests, small differences in large data sets can yield impressively small $p$-values and thus claims of \emph{statistical significance} even when the difference is irrelevant in practice and the significance is statistical only~\cite{Coffin00}.
Using confidence intervals (Section~\ref{sec:stats:confidence-intervals}) or the posterior distribution in addition with a region of practical equivalence avoids this problem~\cite{kruschke2018bayesian}.

\begin{table}
\centering
\begin{tabular}{cccc}
 & & Frequentist & Bayesian\\
 \toprule
 \multicolumn{2}{c}{Estimation} & Confidence Interval & Posterior\\
 \midrule
 \multirow{3}{*}{Hypothesis} & Equivalence & TOST & \multirow{2}{*}{Posterior + ROPE \emph{or} BF} \\
 \cline{2-3}
 & Difference & NHST &\\
 \cline{2-4}
 & Model Selection & AIC & Bayes Factor (BF)\\
\end{tabular}
\caption{Overview of different statistical methods.
 TOST stands for \emph{Two One-Sided T-Tests} and AIC is the \emph{Akaike Information Criterion}~\cite{akaike1974statistical}, a combined measure for model fit and model complexity.
 }
\label{table:stats-methods}
\end{table}

\begin{figure}
\tikzset{%
   >={Latex[width=2mm,length=2mm]},
  base/.style = {rectangle, draw=black, minimum width=2.5cm, minimum height=1cm, text centered},
  inner/.style = {base, rounded corners},
  leaf/.style = {base, fill=blue!15},
}
 \begin{tikzpicture}[node distance=4.35cm,scale=0.8,
    every node/.style={fill=white}, align=center]
  \node (root) [inner] {Hypothesis test or Parameter estimation?};
  \node (hypotheses) [inner,below right of=root] {Difference,\\ Equivalence\\ or Model Comparison?};
  \node (difference) [inner,below right of=hypotheses] {Hypothesis of difference\\ possible to model?};
  \node (nhst) [leaf,below right of=difference] {Null Hypothesis\\ Statistical Tests\\ \ref{sec:stats:nhst}};
  \node (rope) [leaf,below left of=difference] {Region of\\ Practical Equivalence\\ \ref{sec:stats:equivalence}};
  \node (bf) [leaf,above right of=difference] {Bayes Factor\\ \ref{sec:stats:bayes-factor}};
  \node (parameter) [inner,below left of=root] {Bayesian or Frequentist\\ preference?};
  \node (credible) [leaf,below right of=parameter] {Credible Interval\\ \ref{sec:stats:bayesian-inference}};
  \node (confidence) [leaf, below left of=credible] {Confidence Interval\\ \ref{sec:stats:confidence-intervals}};
  
  \draw[->] (root) -- node {Hypothesis} (hypotheses);
  \draw[->] (root) -- node {Parameter} (parameter);
  \draw[->] (parameter) -- node {Bayesian} (credible);
  \draw[->] (parameter) -- node {Frequentist} (confidence);
  \draw[->] (hypotheses) -- node {Model\\Comparison} (bf);
  \draw[->] (hypotheses) -- node {Difference} (difference);
  \draw[->] (hypotheses) -- node {Equivalence} (rope);
  
  \draw[->] (difference) -- node {No} (nhst);
  \draw[->] (difference) -- node {Yes} (rope);
  
 \end{tikzpicture}
 \caption{Flowchart outlining our method selection.}
 \label{fig:stats-flowchart}
\end{figure}
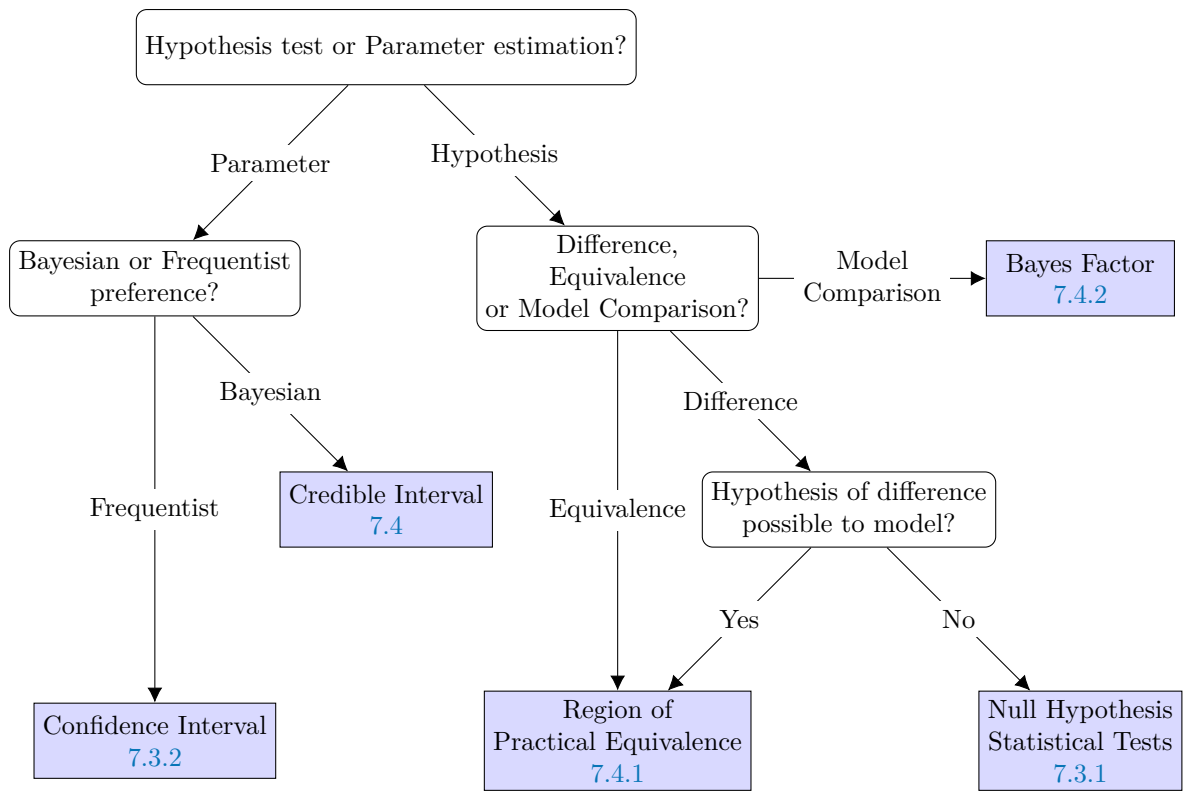

Below is a rough guideline (also shown in Figure~\ref{fig:stats-flowchart}) outlining our method selection process.
It favors Bayesian methods, since the python library PyMC3~\cite{salvatier2016probabilistic} offering them fits well into our workflow.
\begin{enumerate}
\item Define a model that captures the parts of the measured results that interest you, see Section~\ref{sec:stats:model}.
\item Using \emph{confidence intervals} (Section~\ref{sec:stats:confidence-intervals}) or \emph{credible intervals} (Section~\ref{sec:stats:bayesian-inference}), estimate plausible values for the model parameters, including their uncertainty.
If this proves intractable and you are only interested in whether a measured difference is due to chance, use a significance test instead (Section~\ref{sec:stats:nhst}).
\item If you want to show that two distributions (of outcomes of algorithms) are \emph{similar}, use an equivalence test, in which you define a region of practical equivalence.
\item If you want to show that two distributions (of outcomes of algorithms) are \emph{different}, you may also use an equivalence test or alternatively, a significance test.
\item If you want to compare how well different hypotheses explain the data, for example compare whether the diameter has an influence on relative scaling, compare the \emph{relative fit} using a Bayes factor (Section~\ref{sec:stats:bayes-factor}).
\end{enumerate}

Needless to say, these are only recommendations.

\section{Conclusions}
\label{sec:concl}
Besides setting guidelines for experimental algorithmics, this paper provides a tool for simplifying the
typical experimental workflow. We hope that both are useful for the envisioned target group -- and beyond, of course.
We may have omitted material some consider important.
This happened on purpose to keep the exposition reasonably concise.
To account for future demands, we could imagine an evolving \quot{community version} of the paper, updated 
with the help of new co-authors in regular time intervals.
That is why we invite the community to contribute comments, corrections and/or
text.\footnote{The source files of this paper can be found
	at \url{https://github.com/hu-macsy/ae-tutorial-paper}. We encourage
	readers post suggestions via GitHub issues and welcome pull requests.}

Let us conclude by reminding the reader: most of the guidelines in the paper are not scientific laws nor set in stone.
Always apply common sense to adapt a guideline to your concrete situation!
Also, we cannot claim that we always followed the guidelines in the past.
But this was all the more motivation for us to write this paper, to develop \exptool 
and to have a standard to follow -- we hope that the community shares this motivation.

\acknowledgments{We thank Michael Hamann and Sebastian Schlag for their timely and
  thorough feedback on a working draft of the paper. We also thank Alexander Meier and Dimitri
  Ghouse for valuable advice on statistics. A subset of the authors was partially supported
  by grant ME 3619/3-2 within German Research Foundation (DFG)
  Priority Programme 1736 Algorithms for Big Data.}

\bibliographystyle{nws}
\nocite{*}
\bibliography{biblio,datavis}

\appendix

\end{document}